\documentclass[%
 amsmath,amssymb,
 reprint,%
]{revtex4-2}
\usepackage[colorlinks=true, allcolors=blue]{hyperref}
\usepackage{graphicx}
\usepackage{siunitx}
\usepackage{ulem}
\usepackage{url}
\usepackage{xcolor}
\usepackage{dcolumn}
\usepackage{bm}

\begin{document}

\preprint{AIP/123-QED}

\title[]{Multiscale studies of delayed afterdepolarizations II: Calcium-overload-induced ventricular arrhythmias}

\author{Navneet Roshan}
 \affiliation{Centre for Condensed Matter Theory, Department of Physics, Indian Institute of Science, Bangalore, 560012, India \\}
\author{Rahul Pandit}
\affiliation{Centre for Condensed Matter Theory, Department of Physics, Indian Institute of Science, Bangalore, 560012, India \\}
 \email{rahul@iisc.ac.in}

\date{\today}

\begin{abstract}
Disturbances in calcium homeostasis in a cardiac myocyte can lead to calcium-overload conditions and abnormal calcium releases,
which occur primarily in the following two phases of the action potential (AP): (a) triggered or late calcium release (LCR) during the plateau phase; (b) spontaneous calcium release (SCR) during the diastolic interval (DI). Experimental and numerical studies of LCRs and SCRs have suggested that these abnormal calcium releases can lead to triggered excitations and thence to life-threatening ventricular arrhythmias. We explore this suggestion in detail by building on our work in the previous accompanying Paper I, where we have studied abnormal calcium releases and delayed afterdepolarizations (DADs) in two state-of-the-art mathematical models for human ventricular myocytes. Here, we carry out a detailed \textit{in-silico} study of one of these models, namely, the  ten Tusscher-Panfilov TP06~\cite{ten2006alternans} model. We increase the L-type Ca-channel current $I_{\rm{CaL}}$, to trigger LCRs, and  calcium leak through the ryanodine receptor (RyR), to trigger SCRs, in the myocyte. We then perform multiscale simulations of coupled TP06-model myocytes in tissue in one-, two-, and three-dimensional (1D, 2D, and 3D) domains, with clumps of DAD-capable myocytes, to demonstrate how these clumps precipitate premature ventricular complexes (PVCs) that lead, in turn, to fibrillatory excitations like spiral and scroll waves. We examine possible pharmacological implications of our study for the class of ventricular arrhythmias that result from Ca\textsuperscript{2+} overload. 

\end{abstract}

\maketitle

\section{Introduction}
\label{sec:Intro}
Mammalian hearts are made up of electrically excitable tissue in which individual muscle cells (or cardiomyocytes) are electrically coupled with their neighbors. Normal electrical stimulation starts from the sino-atrial node (SAN) and propagates, in the form of an electrical wave of activation, through this tissue, mediates the synchrony of the myocytes, and results in the normal rhythm of the heart. Cardiac arrhythmias, which arise from disturbances in this normal propagation, are among the leading causes of mortality~\cite{mehra2007global}; and risk factors associated with arrhythmias have become worse after COVID-19~\cite{patone2022risks}. Ventricular arrhythmias, like ventricular tachycardia (VT) and ventricular fibrillation (VF), can lead to sudden cardiac death (SCD). These arrhythmias are associated with reentrant waves such as spiral or scroll waves, which have been studied extensively in mathematical models for cardiac tissue. VT or VF can arise in several ways [see, e.g., Refs.~\cite{antzelevitch2011overview,clayton2011models,niederer2019computational}], some of which originate at  cellular and sub-cellular scales~\cite{keating2001molecular}. One important sub-cellular precursor that can lead to such arrhythmias is a disruption of the Ca\textsuperscript{2+} handling inside the myocyte. \\

The concentration of Ca\textsuperscript{2+} inside a myocyte plays a critical role in cell functions; and it is coupled to the membrane potential $V_{\rm{m}}$, so it can affect the myocyte action potential (AP). The proteins in cardiomyocytes help to maintain multiple compartments, with varying amounts ~\cite{petersen2002calcium} of Ca{\textsuperscript{2+}} concentrations [Ca\textsuperscript{2+} homeostasis]. The disruption in Ca\textsuperscript{2+} homeostasis can occur because of upregulations, 
downregulations, or mutations of proteins involved in calcium handling~\cite{wleklinski2020molecular}. An increase in the intracellular Na\textsuperscript{+} concentration~\cite{pogwizd2003intracellular}, 
the use of certain medications~\cite{hauptman1999digitalis}, ischemia ~\cite{marban1989calcium}, and catecholamines~\cite{priori1990mechanisms,opie1985calcium} also disrupt Ca\textsuperscript{2+} homeostasis. Many of these conditions are associated with a Ca\textsuperscript{2+} overload in the cell or an abnormal release of Ca\textsuperscript{2+} from the sarcoplasmic reticulum (SR) to the cytosol or both [see Fig. 1 of the previous accompanying Paper I (henceforth Paper I)].

The abnormal release of Ca\textsuperscript{2+} occurs primarily in the following two phases or intervals of the AP: (a) triggered or late Ca\textsuperscript{2+} release (LCR) during the plateau phase~\cite{shiferaw2012intracellular,fowler2018late}; (b) spontaneous Ca\textsuperscript{2+} release (SCR) during the diastolic interval (DI)~\cite{bassani1995rate}; some studies have shown that both LCRs and SCRs have been observed in ischemia~\cite{lascano2013role}, in hypertrophied~\cite{wei2010t} or failing hearts~\cite{hoang2015calcium}, and also during catecholaminergic polymorphic ventricular tachycardias (CPVTs)~\cite{wleklinski2020molecular}. Here, abnormal Ca\textsuperscript{2+} releases affect $V_{\rm{m}}$ through the electrogenic sodium-calcium exchanger NCX (or NaCa) 
and thus disturb the phase of 
the AP, in which they occur; the strength of these releases determines the amplitude of the disturbances in $V_{\rm{m}}$. The occurrence of LCRs, during the plateau phase of the AP, promotes the elongation of the action-potential duration (APD) via early afterdepolarizations (EADs). Similarly, SCRs are responsible for delayed afterdepolarizations (DADs) in the AP. Such EADs and DADs can lead to premature ventricular complexes (PVCs)~\cite{wit1983pathophysiologic,janse2004electrophysiological,zimik2015comparative,pogwizd2004cellular}. We refer the reader to Refs.~\cite{song2015calcium,zimik2015comparative,vandersickel2014study} for discussions of the different types of EADs and DADs.

The SCR-driven DADs can lead to premature ventricular complexes (PVCs)~\cite{xie2010so,houser2000does}, if there is synchrony between DADs of neighboring myocytes and the source-sink relationship~\cite{spector2013principles} in cardiac tissue, which occurs, e.g., in non-ischemic heart failures~\cite{pogwizd2004cellular}, hypertrophied failing hearts~\cite{de2000contractile}, the post-acidotic incidence of cardiac arrhythmias~\cite{pogwizd1986induction}, and CPVTs \cite{leenhardt1995catecholaminergic}. These PVCs are one of the causes of arrhythmias~\cite{pogwizd2004cellular}. Numerical simulations have suggested that DAD-driven PVCs can promote 
reentry and spiral waves if (a) subthreshold DADs (defined in Paper I Fig. 5) interact with infarcted cardiac tissue (as shown, e.g., for porcine cardiac tissue in Ref.~\cite{campos2022subthreshold}) or (b) there is an interplay of DADs with mutated fast Na\textsuperscript{+} ion channels (as shown, e.g., for leporine cardiac tissue in Ref.~\cite{liu2015delayed}). Furthermore, DAD-driven PVCs are associated with CPVT, which is responsible for SCDs in young and healthy individuals~\cite{wleklinski2020molecular,leenhardt1995catecholaminergic,liu2008catecholaminergic,priori2002clinical} or sports-related SCDs~\cite{stormholt2021symptoms}; in the latter two cases, hearts may not have infarcted tissue or mutated Na\textsuperscript{+} channels. \\

Recent experiments and a few numerical studies suggest that both LCRs and SCRs can be present in a 
myocyte during Ca\textsuperscript{2+} overload~\cite{volders1997similarities,fowler2018late,fowler2020arrhythmogenic,shiferaw2012intracellular,antzelevitch1995clinical}; these studies shed some light on the arrhythmogenic role of these LCRs, but a detailed mechanistic understanding has to be developed.

It is important, therefore, to study DAD-driven arrhythmias by using state-of-the-art computational models and methods, which are becoming increasingly important in the study of all cardiac arrhythmias~\cite{clayton2011models,niederer2019computational}.

Our study uses the ten Tusscher-Panfilov TP06~\cite{ten2006alternans} mathematical model to explore whether Ca\textsuperscript{2+}-release events, such as LCRs and SCRs, can promote reentry in ventricular tissue via the formation of PVCs.
To the best of our knowledge there has been no comprehensive study, at the tissue level, of ventricular arrhythmias induced by a group of myocytes that are capable of abnormal calcium release and which can lead to PVCs in tissue with no infarction-induced scars.

\begin{figure}
    \centerline{\includegraphics [width=0.95\linewidth] {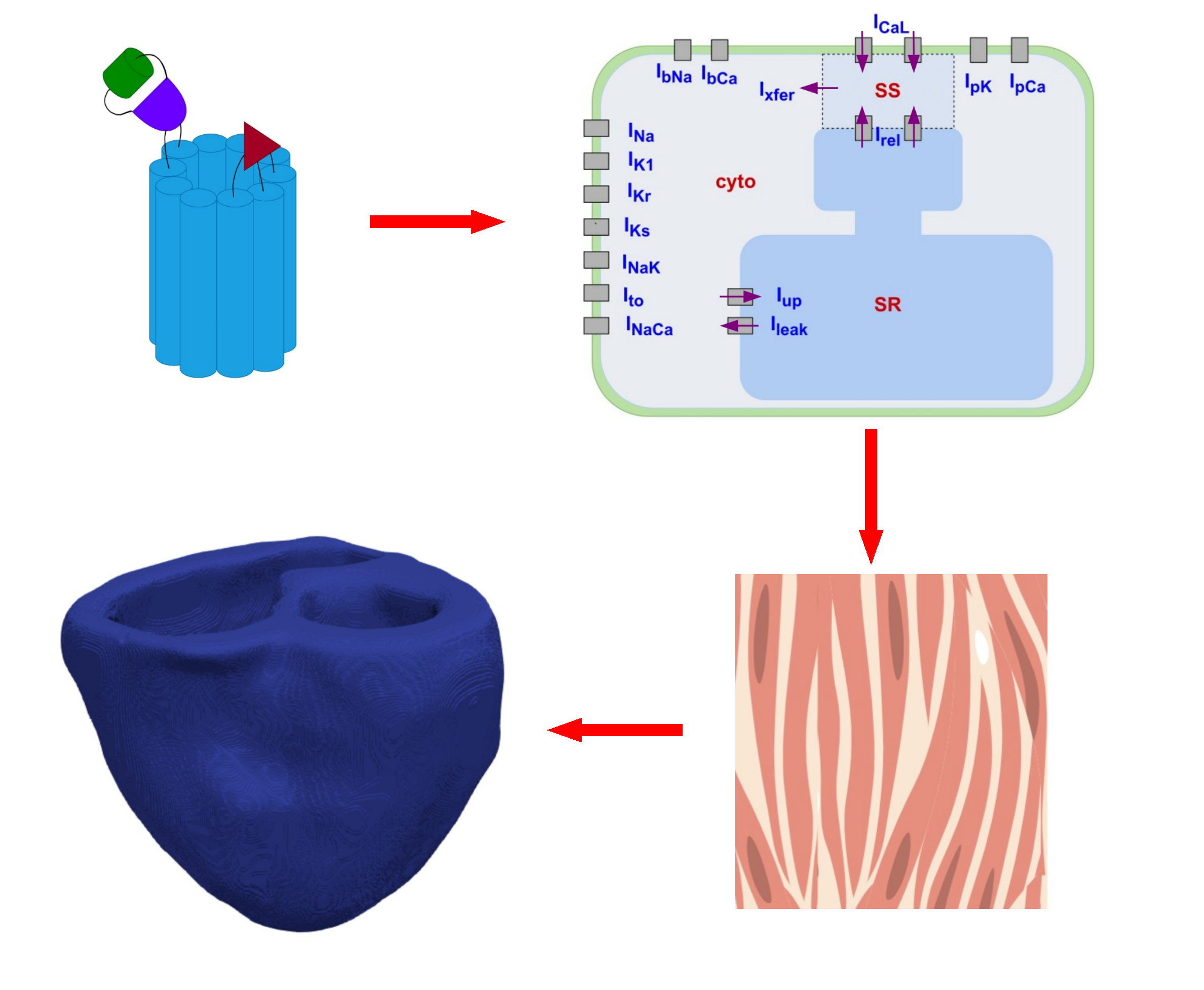}}
    \caption{(Color online) Schematic diagram illustrating the multi-scale nature of ventricular-tissue modeling: at small scales, there are (top left) ion channels, pumps, or exchangers, bounded by proteins, which are shown schematically as blue cylinders, and regulatory molecules, indicted by a red arrowhead and a blue-green doublet; a cardiac myocyte (an excitable cell)  has many such channels, pumps, and exchangers that are shown schematically in the top-right panel for the TP06 model (see Tab.~\ref{tab:tp_currents} and Fig. 1 of the accompanying Paper I); cardiac tissue, depicted schematically in the bottom-right picture [image made via the software Mind the Graph (see \url{https://mindthegraph.com})], contains many such myocytes, oriented in a particular direction, and other cells such as fibroblasts (see, e.g., Ref.~\cite{nayak2013spiral}) and Purkinje fibers (see, e.g., Ref.~\cite{nayak2017spiral}); human ventricles and atria are made up of such cardiac tissue (at the bottom left we present an anatomically realistic human bi-ventricular domain).}
    \label{fig:multiscale}
\end{figure}

We provide a theoretical framework for investigating ventricular arrhythmias that arise from Ca\textsuperscript{2+} overload [see, e.g., Refs. ~\cite{tiso2001identification,janse2004electrophysiological,hove2004atrial,coppini2018altered}] and which are associated with LCR- and SCR-induced SCDs. Only a few mathematical models for ventricular tissue can yield both LCRs and SCRs 
[see, e.g., Paper I and Ref.~\cite{fink2011ca2+}]. We use one of these models, namely, the TP06 model, because it shows various types of DADs at the myocyte scale. Our multi-scale study, which spans myocyte, cable, tissue, and ventricular scales [see the schematic diagram in Fig.~\ref{fig:multiscale}], yields several new insights into PVC-induced \textit{non-shockable arrhythmias}~\cite{fan2012shockable} that are precipitated by Ca\textsuperscript{2+} overload and which cannot be eliminated by straightforward low-amplitude electrical-defibrillation protocols.

  Before we present the details of our study, we give a qualitative overview of our principal results. In Subsection~\ref{subsec:protocol}, we increase $I_{\rm{CaL}}$ to provide the Ca\textsuperscript{2+}-overload and we quantify its effects on $V_{\rm{m}}$ and the Ca\textsuperscript{2+} concentration in the sarcoplasmic store, for two Ca\textsuperscript{2+}-overload protocols. In Subsection~\ref{subsec:afterdep}, we demonstrate the emergence of LCR-induced EADs and SCR-triggered DADs in the TP06 myocyte model. In Subsection~\ref{subsec:lcclate} we show that a selective reduction of the late part of $I_{\rm{CaL}}$ can eliminate LCRs and hence EADs; we distinguish two types of myocytes: those that are capable of producing DADs (henceforth, DAD myocytes) and those that cannot (henceforth, non-DAD myocytes). In the next Subsection~\ref{subsec:cable} we investigate the propagation of electrical waves of activation in a cable-type domain with non-DAD myocytes with a clump of DAD myocytes in the middle; we examine this propagation for the two Ca\textsuperscript{2+}-overload protocols [Subsection~\ref{subsec:protocol}] with the Ca\textsuperscript{2+}-overload either localized to the DAD-clump or spread over the entire cable. We demonstrate, then, that (a) the DAD clump can lead to the emergence of DAD-driven premature ventricular complexes (PVCs) and (b) the interaction of successive propagating PVCs results in conduction blocks. Our simulations in two-dimensional (2D) [Subsection~\ref{subsec:2D}] and three-dimensional (3D) anatomically realistic human bi-ventricular [Subsection~\ref{subsec:3D}] domains, with DAD clumps, also display such PVCs that lead to reentry, i.e., the formation of spiral (2D) or scroll (3D) waves. We show explicitly that these PVCs cannot be removed by a standard low-amplitude electrical-defibrillation technique.

 We have organized the rest of this paper as follows. In Sec.~\ref{sec:Mat-and-Methods}, we describe the models we use and the numerical and theoretical methods that we employ. Section~\ref{sec:results} is devoted to a detailed discussion of our results. The concluding Section~\ref{sec:disc_and_concl} examines the possible clinical implications of our principal results.

\section{Model and Methods}{\label{sec:Mat-and-Methods}}
In Paper I, we have studied two human-ventricular mathematical models. Here, we consider one of them, namely, the TP06 model. We stimulate the TP06-model myocytes by an electric current (square pulses of height $-52 ~\text{pA/pF}$  and duration $1 ~\text{ms}$ with a 1 Hz pacing frequency). 

\subsection{TP06 Model}
\label{section:model}
In the TP06 model~\cite{ten2006alternans} for cardiac tissue, we use the following partial differential equation (PDE) for the transmembrane potential $V_{\rm{m}}$:

\begin{align}
\label{eq:TP06pde}
 \dfrac{\partial V_{\rm{m}}}{\partial t} = & -\dfrac{I_{\rm{stim}} + I_{\rm{TP06}}}{C_{\rm{m}}} + \nabla. \left( \mathbf{D} \nabla V_{\rm{m}}\right)\,;
\end{align}
$t$ is the time; $C_{\rm{m}}$ is the capacitance per unit area of the myocyte membrane; $I_{\rm{stim}}$ is the externally applied current stimulus to the myocyte; $I_{\rm{TP06}}$ is the sum of all the transmembrane ionic currents [Eq.~\ref{eq:TP06currents}]; and $\mathbf{D}$ is the diffusion tensor, which is taken to be a scalar for cable (1D) and two-dimensional (2D) simulation domains
and a tensor for the 3D bi-ventricular domain; in isolated-myocyte studies, there is no diffusion term. The single-myocyte TP06 model contains the following 12 currents that contribute to the dynamics of $V_{\rm{m}}$:
\begin{eqnarray}
I_{\rm{TP06}}&=&I_{\rm{Na}}+I_{\rm{CaL}}+I_{\rm{K1}}+I_{\rm{Kr}}+I_{\rm{Ks}}+I_{\rm{to}}+I_{\rm{pK}} \nonumber  \\
&+&I_{\rm{bCa}}+I_{\rm{NaCa}}+I_{\rm{NaK}}+I_{\rm{bNa}}+I_{\rm{pCa}}.
\label{eq:TP06currents}
\end{eqnarray}

\begin{table}[ht]
	\centering 
	\begin{tabular}{|c|c|} 
\hline	
 $I_{\rm{Na}}$ & fast inward Na\textsuperscript{+} current \\
\hline
 $I_{\rm{CaL}}$ & L-type Ca\textsuperscript{2+} current \\
\hline
 $I_{\rm{K1}}$ & inward rectifier K\textsuperscript{+} current \\
\hline 
$I_{\rm{Kr}}$ & rapid delayed rectifier K\textsuperscript{+} current \\
\hline
 $I_{\rm{Ks}}$ & slow delayed rectifier K\textsuperscript{+} current \\
\hline
 $I_{\rm{to}}$ & transient outward K\textsuperscript{+} current \\
\hline
 $I_{\rm{pK}}$ & plateau  K\textsuperscript{+} current \\
\hline
 $I_{\rm{bCa}}$ & background Ca\textsuperscript{2+} current \\
\hline
 $I_{\rm{NaCa}}$ & Na\textsuperscript{+}-Ca\textsuperscript{2+} exchanger current \\
\hline
 $I_{\rm{NaK}}$ & Na\textsuperscript{+}- K\textsuperscript{+} ATPase current \\
\hline
 $I_{\rm{bNa}}$ & Na\textsuperscript{+} background current \\
\hline
 $I_{\rm{pCa}}$ & plateau Ca\textsuperscript{2+} current \\
 \hline
\end{tabular}
\caption{The ionic currents used in the TP06 model for human-ventricular myocytes; Ref.~\cite{ten2006alternans}  describes the detailed dynamics of these currents.}
\label{tab:tp_currents}
\end{table}
These currents are defined in Table~\ref{tab:tp_currents}; and the full set of ODEs for this model is given in Ref.~\cite{ten2006alternans}. We concentrate on DADs, which are affected significantly by $I_{\rm{CaL}}$, so we give the equation for this current below:

\begin{align}
\label{eq:TP06LCC}
I_{\rm{CaL}}= G_{\rm{CaL}} \cdot d \cdot f \cdot f_2\cdot f_{\rm{Cass}} \cdot F\left( Ca_{\rm{SS}}, V_{\rm{m}}, Ca_{\rm{o}} \right),
\end{align}
where $d$ is a voltage-dependent activation gate, $f$ and $f_{\rm{2}}$ are, respectively, voltage-dependent, slow- and fast-inactivation gates, and $f_{\rm{Cass}}$ is the cytosolic-calcium-dependent inactivation gate; $F$ is a function of the  subspace Ca\textsuperscript{2+} concentration $Ca_{\rm{SS}}$ , $V_{\rm{m}}$, and the extracellular Ca\textsuperscript{2+} concentration $Ca_{\rm{o}}$. $f_{2}$
obeys the following equations:
\begin{eqnarray}
\label{eq:f2}
\dfrac{df_{\rm{2}}}{dt} &=& \dfrac{f_{\rm{2}}-f_{\rm{2}{,\infty}}}{\tau_{f_{\rm{2}}}}\,;  \\
\label{eq:f2_inf}
f_{\rm{2}{,\infty}} &=& \frac{0.67}{1+\exp{\dfrac{V_{\rm{m}}+35}{7}}}+ f_{\rm{{2}{,sat}}}\,; 
\end{eqnarray}
\begin{multline}
\tau_{f_{\rm{2}}} = 562\cdot\exp\dfrac{-\left(V_{\rm{m}}+27\right)^2}{240}+ 
  \frac{31}{1+{\exp\frac{25-V_{\rm{m}}}{10}}}+ \\
  \frac{80}{1+{\exp\frac{V_{\rm{m}}+30}{10}}}\,;
\end{multline}
here, $f_{\rm{2}{,sat}}$ determines the late phase of $I_{\rm{CaL}}$ and $\tau_{f_{\rm{2}}}$ is a time constant; the control value of $f_{\rm{2}{,sat}} = 0.33$ in the TP06 model. In  Subsection~\ref{subsec:lcclate}, we tune the value of $f_{\rm{2}{,sat}}$ to suppress LCRs.
The following currents (see Paper I) also play an important role in these LCRs:
\begin{eqnarray}
\label{eq:serca}
I_{\rm{up}} &=& {\frac {V_{\rm{maxup}}}{1+{\left ({\frac {K_{\rm{up}}}{{Ca_{\rm{i}}}}}\right )}^2}};\\
I_{\rm{leak}} &=& V_{\rm{leak}}\cdot(Ca_{\rm{SR}}- Ca_{\rm{i}});
\end{eqnarray}
here, $I_{\rm{up}}$, $K_{\rm{up}}$, and $V_{\rm{maxup}}$ are, respectively, the sarco-endoplasmic-reticulum Ca\textsuperscript{2+} ATPase (SERCA) uptake rate, a constant, and the maximal SERCA uptake rate; $Ca_{\rm{i}}$ is the cytosolic calcium concentration; $I_{\rm{leak}}$ is the 
leak rate; $V_{\rm{leak}}$ dictates the strength of the leakage; and $Ca_{\rm{SR}}$ is the sarcoplasmic-reticulum (SR) calcium concentration. In our 2D and 3D studies with the TP06 model, we set $I_{\rm{leak}} = 0$.  The opening of the ryanodine receptor (RyR) and calcium release through it is modeled by the following equation: 

\begin{align}
\label{eq:rel}
    I_{\rm{rel}} = (V_{\rm{rel}}.O + V_{\rm{RyRL}})\cdot(Ca_{\rm{SR}}- Ca_{\rm{SS}}),
\end{align}
 where $I_{\rm{rel}}$ is the release rate; $V_{\rm{rel}}$ dictates the strength of the release; $O$, the probability that the RyR is open, is a function of $Ca_{\rm{SR}}$, $Ca_{\rm{SS}}$ (the subspace Ca\textsuperscript{2+} concentration), and other parameters in the TP06 model (see the Appendix of Paper I);  $V_{\rm{RyRL}}$ controls the RyR leak (we use $V_{\rm{RyRL}} = 0$, if there is no RyR leak, and $V_{\rm{RyRL}} = 0.00018 \rm{ms}^{-1}$ to 
introduce a representative RyR leak); we use this 
$V_{\rm{RyRL}}$ to model the calcium leak from the SR to the cytosol which in turn leads to DADs~\cite{chelu2007sarcoplasmic}. We employ the following two ways of increasing the calcium overload for such myocytes: (a) increasing $G_{\rm{CaL}}$; and  (b) increasing $G_{\rm{CaL}}$ along with $G_{\rm{Kr}}$ to maintain the APD. We have discussed (b) in Paper I, which we also follow to scale the maximal conductances or fluxes as in Ref.~\cite{mulimani2022silico}; e.g., to scale $G_{\rm{CaL}}$, we define $G_{\rm{CaL}} \equiv S_{\rm{GCaL}} \times G_{\rm{CaL0}}$, where $G_{\rm{CaL0}}$ is the control value of $G_{\rm{CaL}}$ and $S_{\rm{GCaL}}$ is the scale factor for $G_{\rm{CaL}}$.

\subsection{Numerical Methods}
\label{subsec:nummeth}
	The numerical methods we use are the same as those in Paper I. In short, we use the forward-Euler method for time marching in Eq.~\ref{eq:TP06pde} and the Rush-Larsen scheme for integrating the ODEs for the gating variables in the ionic currents [see, e.g., Eq.~\ref{eq:TP06currents}]. For spatial discretization, we use three-, five-, and seven-point stencils for the Laplacian in one (1D), two (2D), and three dimensions (3D). We apply no flux boundary conditions
 at the boundaries of these simulation domains. The time step $\delta t = 0.02$ \si{ms}; the spatial resolution is $\delta x = 0.025$ \si{cm};and $D = 0.00154$ \si{\centi\meter\squared\per\second}. We have checked that, with these parameters, (a) we obtain a conduction velocity
	$\simeq 68$ \si{\centi\meter\per\second}, which is in the bio-physically relevant range, and (b) the von-Neumann stability criterion is satisfied.
	In our anatomically realistic (AR) simulation, we use the human bi-ventricular geometry and fiber-orientation data [obtained from diffusion tensor magnetic resonance imaging (DTMRI) as in Ref.~\cite{winslow2011cardiovascular}]. The diffusion constant along the fiber is $D_{\parallel}=0.00154$ \si{\centi\meter\squared\per\second} and in the transverse direction $D_{\perp}= D_{\parallel}/4$ \si{\centi\meter\squared\per\second}. In tensorial notation 

 \begin{align} 
     \mathbf{D} = D_{\parallel} \delta_{ij} + \left(D_{\parallel}-D_{\perp}\right)\alpha_i\alpha_j, 
 \end{align}
 where $\delta_{ij}$ is Kronecker delta, and $\alpha_i$, with $i = 1, 2, 3$, are the direction cosines that represent the local orientation of a myocyte. We employ the phase-field method [see, e.g., Refs.~\cite{fenton2005modeling,majumder2016scroll,rajany2021effects,rajany2022spiral}], which eliminates the need for applying the von-Neumann boundary conditions on irregular boundaries.

\section{Results}
\label{sec:results}

We present our results in the following Subsections: In Subsection~\ref{subsec:protocol} we examine the dependence of $V_{\rm{m}}$ and $Ca_{\rm{SR}}$ on the Ca\textsuperscript{2+}-overload protocols that we use. In Subsection~\ref{subsec:afterdep} we demonstrate the emergence of Ca{\textsuperscript{2+}} sparks that are responsible for EADs and DADs.
In Subsection~\ref{subsec:lcclate} we show that a reduction in the late part of the L-type Calcium current (LCC) eliminates
EADs. In Subsection~\ref{subsec:cable} we present the results of our cable simulations in which we introduce a clump of
myocytes that yield DADs. Subsection~\ref{subsec:2D} contains illustrative results for the evolution of DAD-induced PVCs and their development into 
spiral waves; we also examine the dependence of these PVCs on the shape of the DAD clump; and we show that these PVCs are non-shockable. 
In Subsection~\ref{subsec:3D} we generalize our 2D study to one in a 3D anatomically realistic human-bi-ventricular domain.

\subsection{The dependence of $V_{\rm{m}}$ and $Ca_{\rm{SR}}$ on $I_{\rm{CaL}}$}\label{subsec:protocol}
 
\begin{figure}
    \centerline{\includegraphics [width=\linewidth] {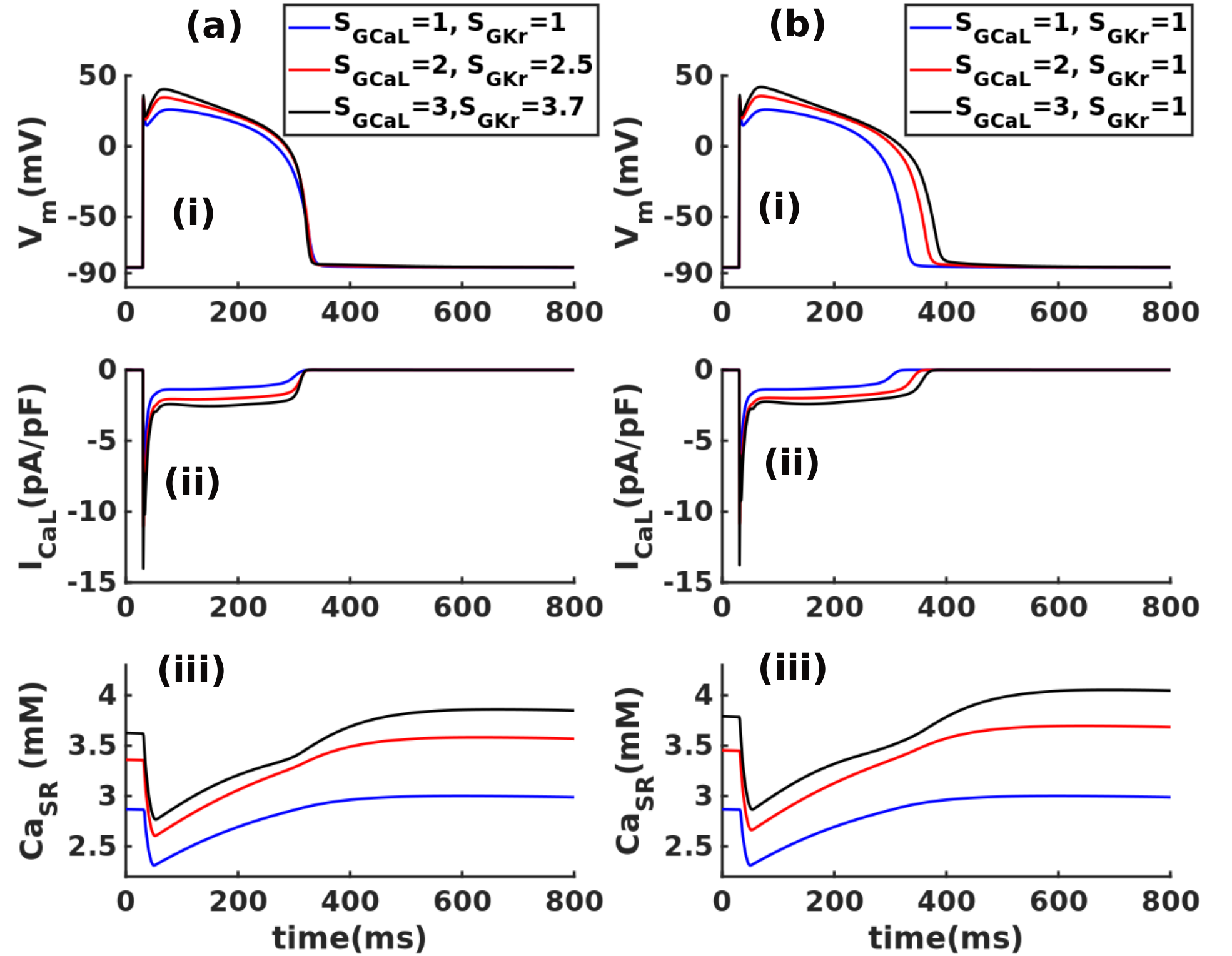}}
    \caption{(Color online) \label{fig:protocol} Plots in panels (a) and (b) compare the two Ca-overload protocols (a) and (b) [see text], for various values of the scale factors $S_{\rm{GCaL}}$ and $S_{\rm{GKr}}$. The subplots (i), (ii), and (iii) show, respectively, how $V_m$, $I_{\rm{CaL}}$, and $Ca_{\rm{SR}}$ change with time $t$.}
\end{figure}

In Figs.~\ref{fig:protocol} (a) and (b) we present plots that compare the two Ca\textsuperscript{2+}-overload protocols, (a) and (b), that we use. 
In protocol (a), we scale the maximal conductance $G_{\rm{CaL}}$ of the L-type calcium channel (LCC) and the maximal conductance $G_{\rm{Kr}}$ of the rapid delayed rectifier potassium channel so that the APD is unchanged. By contrast, in protocol (b), we scale $G_{\rm{CaL}}$ to obtain calcium overload in the myocyte; in this case, the increase in the inward current $I_{\rm{CaL}}$ enhances the APD slightly. In Figs.~\ref{fig:protocol} (a) and (b) the subplots (i), (ii), and (iii) show, respectively, how $V_m$, $I_{\rm{CaL}}$, and $Ca_{\rm{SR}}$ change with time $t$; we have recorded the data for these plots after the first pacing.

In summary, in protocol (a), we scale the $G_{\rm{CaL}}$ and $G_{\rm{Kr}}$ together to increase the $Ca_{\rm{SR}}$, \textit{without changing the APD}; by contrast, in protocol (b), the APD and $Ca_{\rm{SR}}$ \textit{both increase concomitantly} as we increase $G_{\rm{CaL}}$. By comparing subplots (i) and (ii) we see that an increase in APD elongates $I_{\rm{CaL}}$ too. From subplot (iii), we conclude that the SR calcium content $Ca_{\rm{SR}}$ increases with $S_{\rm{GCaL}}$ and it is  higher in protocol (b) than in protocol (a).

\subsection{Emergence of Ca{\textsuperscript{2+}} sparks and afterdepolarizations}
\label{subsec:afterdep}
\begin{figure}
    \centerline{\includegraphics [width=\linewidth] {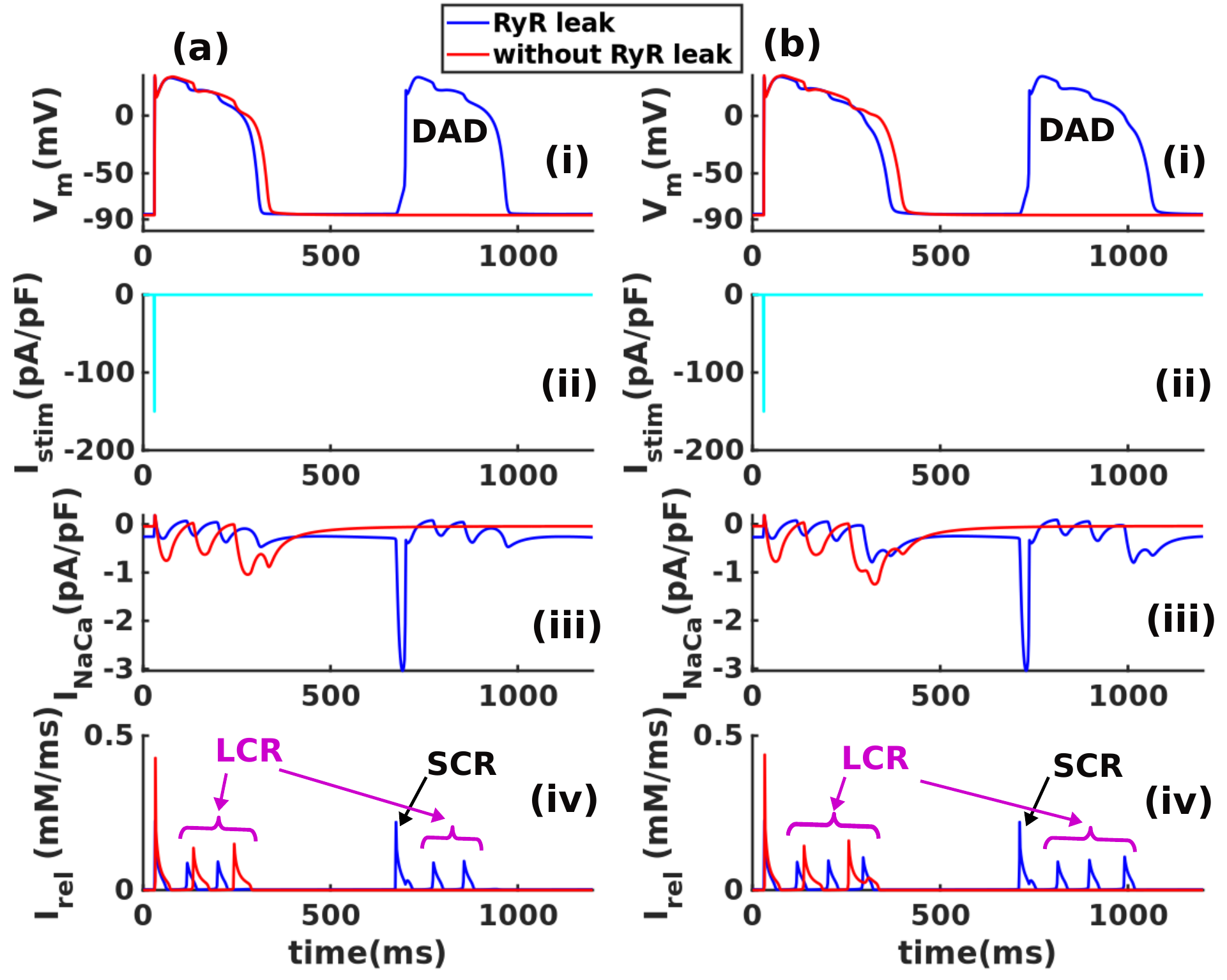}}
    \caption{(Color online) \label{fig:pacing_effect} Plots in the panels (a) and (b) present the effects of pacing (after 15 pacing stimulations) for the two protocols of calcium overload, for two cases, with an RyR leak (in blue) and without an RyR leak (in red). Subplot (i): $V_{\rm{m}}$ for both the cases (with an RyR leak and without an RyR leak); DADs emerge with the RyR leak in both protocols; subplot (ii): current stimulus $I_{\rm{stim}}$; subplot (iii): $I_{\rm{NaCa}}$; subplot (iv): Ca\textsuperscript{2+} release flux, i.e., $I_{\rm{rel}}$ from the RyR.}
\end{figure}

To explore the formation of Ca{\textsuperscript{2+}} sparks, we stimulate the myocyte by using the calcium-overload protocols (a) and (b) [see Subsection~\ref{subsec:protocol}], with the following illustrative parameter values for these two protocols: 
(a) $S_{\rm{Vmaxup}}=4.5$, $S_{\rm{KNaCa}}=2$, $S_{\rm{GCaL}} = 2$, $S_{\rm{GKr}} = 2.5$; and (b) $S_{\rm{Vmaxup}}=4.5$, $S_{\rm{KNaCa}}=2$, $S_{\rm{GCaL}} = 2$, $S_{\rm{GKr}} = 1$. For each protocol, we present two sets of results, namely, with an RyR leak (blue curves) and without an RyR leak (red curves) [i.e., $V_{\rm{RyRL}} = 0.00018 \rm{ms}^{-1}$ and $V_{\rm{RyRL}} = 0$, respectively, in Eq.~(\ref{eq:rel})]. In Fig.~\ref{fig:pacing_effect} we show plots versus time $t$ of (i) $V_{\rm{m}}$, (ii) $I_{\rm{stim}}$, (iii) $I_{\rm{NaCa}}$, and (iv) $I_{\rm{rel}}$, after $15$ pacing stimulations, for protocols (a) and (b) in panels (a) and (b), respectively. From subplots (i), we see that, for both protocols, the inclusion of an RyR leak results in a suprathreshold DAD, whose upstroke is not because of the current stimulus $I_{\rm{stim}}$ (subplots (ii)) but it is associated with an inward spike in $I_{\rm{NaCa}}$ (subplots (iii)) and a sharp upstroke (SCR) in $I_{\rm{rel}}$ (subplots (iv)). [The relation of SCR, $I_{\rm{NaCa}}$, and $V_{\rm{m}}$ has been explained in Paper I. Experiments suggest that ``The SR Ca\textsuperscript{2+} leak through the RyR channel is believed to be the primary mechanisms for SCRs~\cite{wleklinski2020molecular}. The SCRs increase $V_{\rm{m}}$ by activating the Na\textsuperscript{+}/Ca\textsuperscript{2+} exchangers (NCX or NaCa)~\cite{matsuda1997na+} in the forward mode, in which the electrogenic NCX extrudes one Ca\textsuperscript{2+} out of a myocyte and exchanges it with three Na\textsuperscript{+} ions.''] \\
For both protocols (a) and (b), we find oscillations during the plateau phase of the AP (subplot (i)), which are associated with oscillations in $I_{\rm{NaCa}}$ (subplot (ii)) and sharp structures in $I_{\rm{rel}}$ (subplot (iii)). These structures are arising during Ca\textsuperscript{2+} overload and are due to late Ca\textsuperscript{2+} releases (LCRs). With the RyR leak, we observe both LCRs and SCRs. The latter lead to DADs.
Note that the large inward spike in $I_{\rm{NaCa}}$ is associated with the SCR through cytosolic $Ca_{\rm{i}}$ [cf., Eq. (5), in the Appendix of the previous paper, which 
shows the thermodynamic forces on $I_{\rm{NaCa}}$ that arise, \textit{inter alia}, from a competition between inward and outward Na\textsuperscript{+} and Ca\textsuperscript{2+} and also depend on $V_{\rm{m}}$]; during the resting phase of the membrane potential ($V_{\rm{m}}=-86$ mV), the thermodynamic force on $I_{\rm{NaCa}}$ is maximal and its direction is inward~\cite{janvier1996role}, which has the important consequence of unloading Ca\textsuperscript{2+} in the diastolic phase of the AP.   

\subsection{Role of the late part of the $I_{\rm{CaL}}$ current in triggering LCRs}
\label{subsec:lcclate}

\begin{figure}
    \centerline{\includegraphics [width=\linewidth] {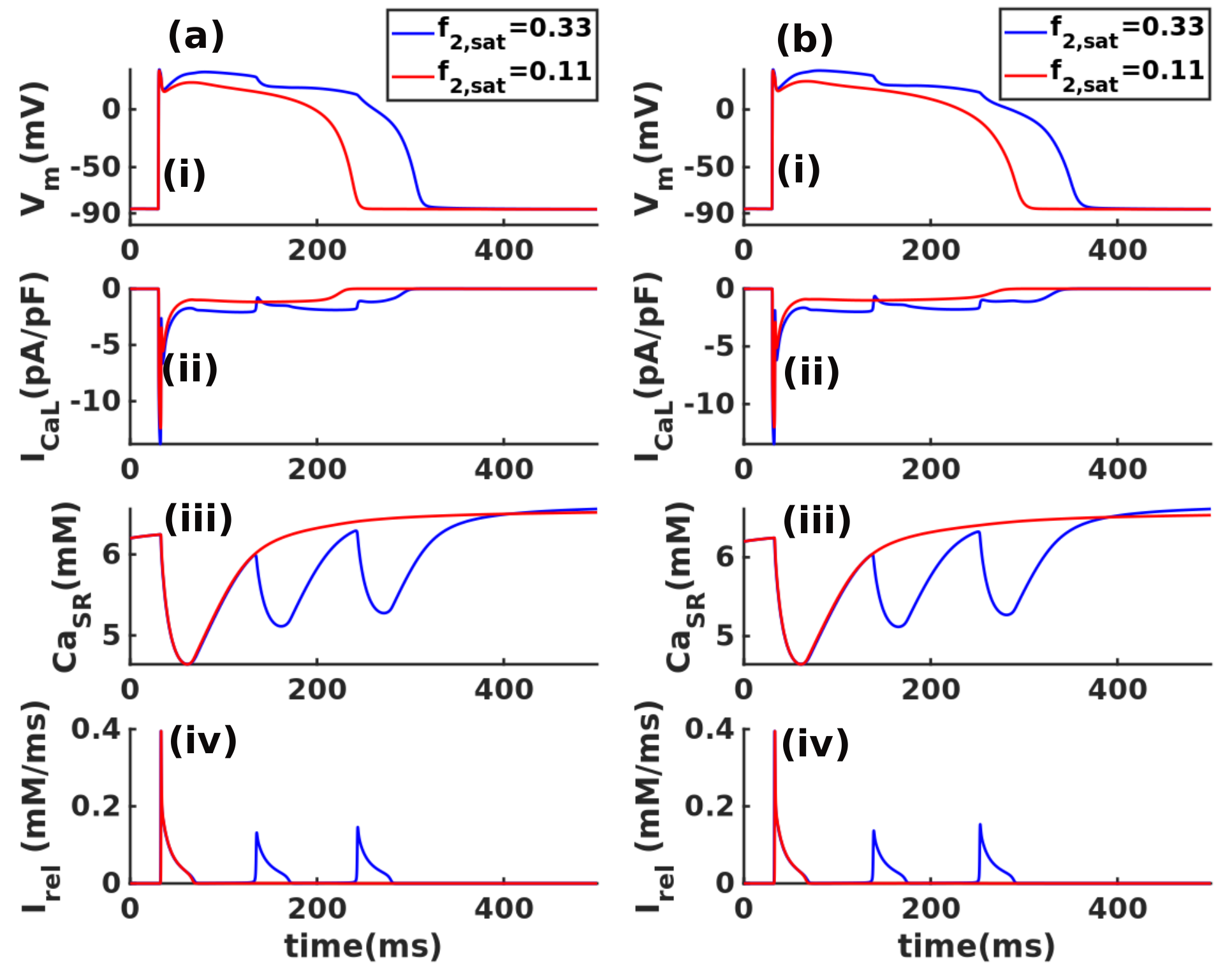}}

    \caption{(Color online) \label{fig:f_sat} Plots versus time $t$ of (i) $V_{\rm{m}}$, (ii) $I_{\rm{CaL}}$, (iii) $Ca_{\rm{SR}}$, and (iv) $I_{\rm{rel}}$, after $15$ pacing stimulations, for protocols (a) and (b) in panels  (a) and (b), respectively.}
\end{figure}

    In the previous Subsection~\ref{subsec:afterdep}, we have shown that, to observe DADs, we need a Ca\textsuperscript{2+} overload and the RyR leak from SR to SS. However, in the model, we have found that LCRs are present even without this RyR leak. 
    Therefore, we now investigate the causes of these LCRs. In the TP06 model, Ca-induced-Ca-release (CICR) is dependent on $Ca_{\rm{SR}}$ and the $Ca_{\rm{SS}}$. The former modulates $O$, the opening probability of the RyR [Eqs.~(\ref{eq:rel}) and (10) in the Appendix of the accompanying paper], whereas the latter acts as a trigger [in Eq. (11) in the Appendix of Paper I]~\cite{fink2011ca2+}. We hypothesize that the late part of the $I_{\rm{CaL}}$ current increases $Ca_{\rm{SS}}$ and is, therefore, responsible for triggering these LCRs; we check this hypothesis as follows:

For the illustrative parameter values $S_{\rm{GCaL}} = 2$, $S_{\rm{GKr}} = 2.5$, $S_{\rm{Vmaxup}}=4.5$, and $S_{\rm{KNaCa}}=2$, we stimulate the myocyte for $15$ pacings, and register the state variables, which we with use as initial conditions. We then stimulate the myocyte, for both protocols (a) and (b), and present the two sets of results, namely, with $f_{\rm{2}{,sat}} = 0.33$, the control value, (blue curves) and $f_{\rm{2}{,sat}} = 0.11$ (red curves) in Eq.~(\ref{eq:f2_inf}), with $V_{\rm{RyRL}} = 0\, \rm{ms}^{-1}$.
In Fig.~\ref{fig:f_sat} we show plots versus the time $t$ of (i) $V_{\rm{m}}$, (ii) $I_{\rm{CaL}}$, (iii) $Ca_{\rm{SR}}$, and (iv) $I_{\rm{rel}}$, for protocols (a) and (b) in panels (a) and (b), respectively. From subplots (i) we see that, for both protocols, the reduction in $f_{\rm{2}{,sat}}$ results in a reduction of the magnitude of $I_{\rm{CaL}}$ (subplot (ii)), in its plateau phase (but not in the initial transient), suppression of oscillations in $Ca_{\rm{SR}}$ (subplot (iii)), and the elimination of the blue spikes in $I_{\rm{rel}}$ (LCRs in subplot (iv)) and a concomitant decrease in the APD. As we have discussed Paper I, the CICR process in the TP06 model depends only on $Ca_{\rm{SR}}$ and $Ca_{\rm{SS}}$; and $Ca_{\rm{SR}}$ is the same for both the cases (a) and (b); therefore, the reduction in $I_{\rm{CaL}}$ is responsible for eliminating LCRs, as $I_{\rm{CaL}}$ directly influences $Ca_{\rm{SS}}$.

It is instructive to compare the EADs discussed in Refs.~\cite{song2015calcium,vandersickel2014study} with the 
LCR-driven EADs (henceforth LCRs) we obtain in Fig.~\ref{fig:pacing_effect}. The former arise because of the reactivation of $I_{\rm{CaL}}$, whereas the latter are associated with the sharp peaks in $I_{\rm{rel}}$, which are triggered by the plateau-part of $I_{\rm{CaL}}$ [see Fig.~\ref{fig:pacing_effect} (a) (iv)].

\subsection{Cable simulations}
\label{subsec:cable}

\begin{figure}
    \centerline{\includegraphics [width=\linewidth] {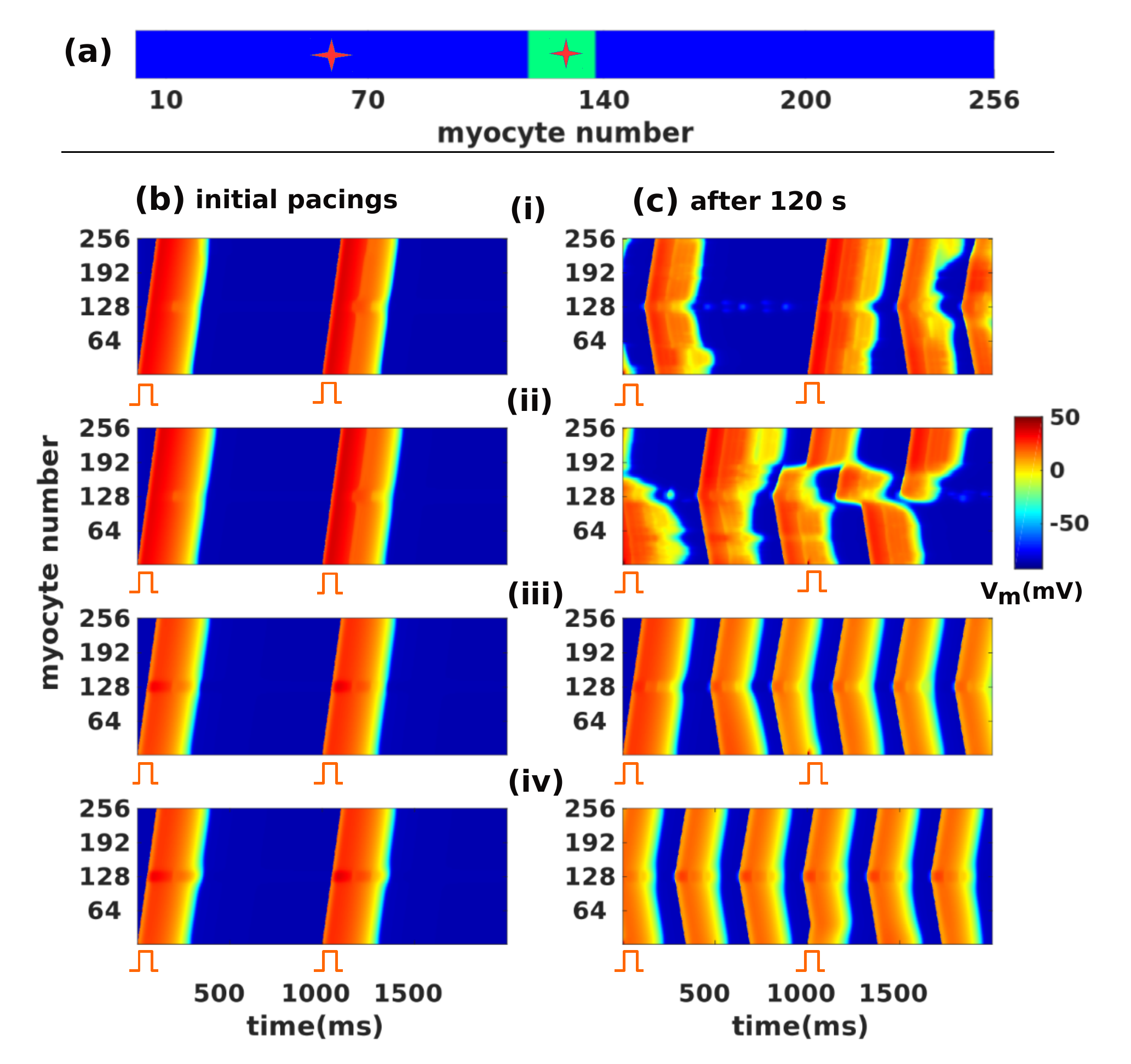}}
    \caption{(Color online) \label{fig:cable_cases} (a) Schematic diagram of our simulation domain for a cable of myocytes: the green region is
    a clump of DAD myocytes (with the RyR leak) and the blue area comprises normal myocytes (without the RyR leak); later we present time series from the representative points marked by red stars. Space-time plots 
    of the membrane potential $V_{\rm{m}}$ for Cases (i)-(iv) [see  Table~\ref{tab:cable_configs} and the text] for (b) the initial two pacings and the (c) after $120$s (the last two pacings).}
\end{figure}

We now carry out simulations in a 1D cable domain, by using the numerical methods of
Subsection~\ref{subsec:nummeth}; this cable has $256$ myocytes, of which $20$ myocytes in the center of the domain are DAD-capable myocytes, which have an RyR leak ($V_{\rm{RyRL}} = 0.00018\, \rm{ms}^{-1}$) and a calcium overload as in Subsection~\ref{subsec:afterdep}. The schematic diagram in Fig. \ref{fig:cable_cases}(a) shows a cable with non-DAD (blue) and the DAD-capable (green) myocytes. We stimulate the first myocyte in the cable and apply the no-flux boundary condition at both ends of the cable. To model the various calcium-overload situations that may arise in pathophysiological conditions, we have considered the four sets of parameters mentioned in Table~\ref{tab:cable_configs}:
\begin{itemize}
    \item Case(i): Ca\textsuperscript{2+} overload in the entire cable (both normal and DAD-capable myocytes); protocol (i) of  Subsection~\ref{subsec:afterdep}.
    \item Case(ii): Ca\textsuperscript{2+} overload in the entire cable (both normal and DAD-capable myocytes); protocol (ii) of  Subsection~\ref{subsec:afterdep}.
    \item Case(iii): Ca\textsuperscript{2+} overload localized at DAD-capable myocytes; protocol (i) of  Subsection~\ref{subsec:afterdep}.
    \item Case(iv): Ca\textsuperscript{2+} overload localized at DAD-capable myocytes; protocol (ii) of  Subsection~\ref{subsec:afterdep}.
\end{itemize} 

\begin{figure}
    \centerline{\includegraphics [width=\linewidth] {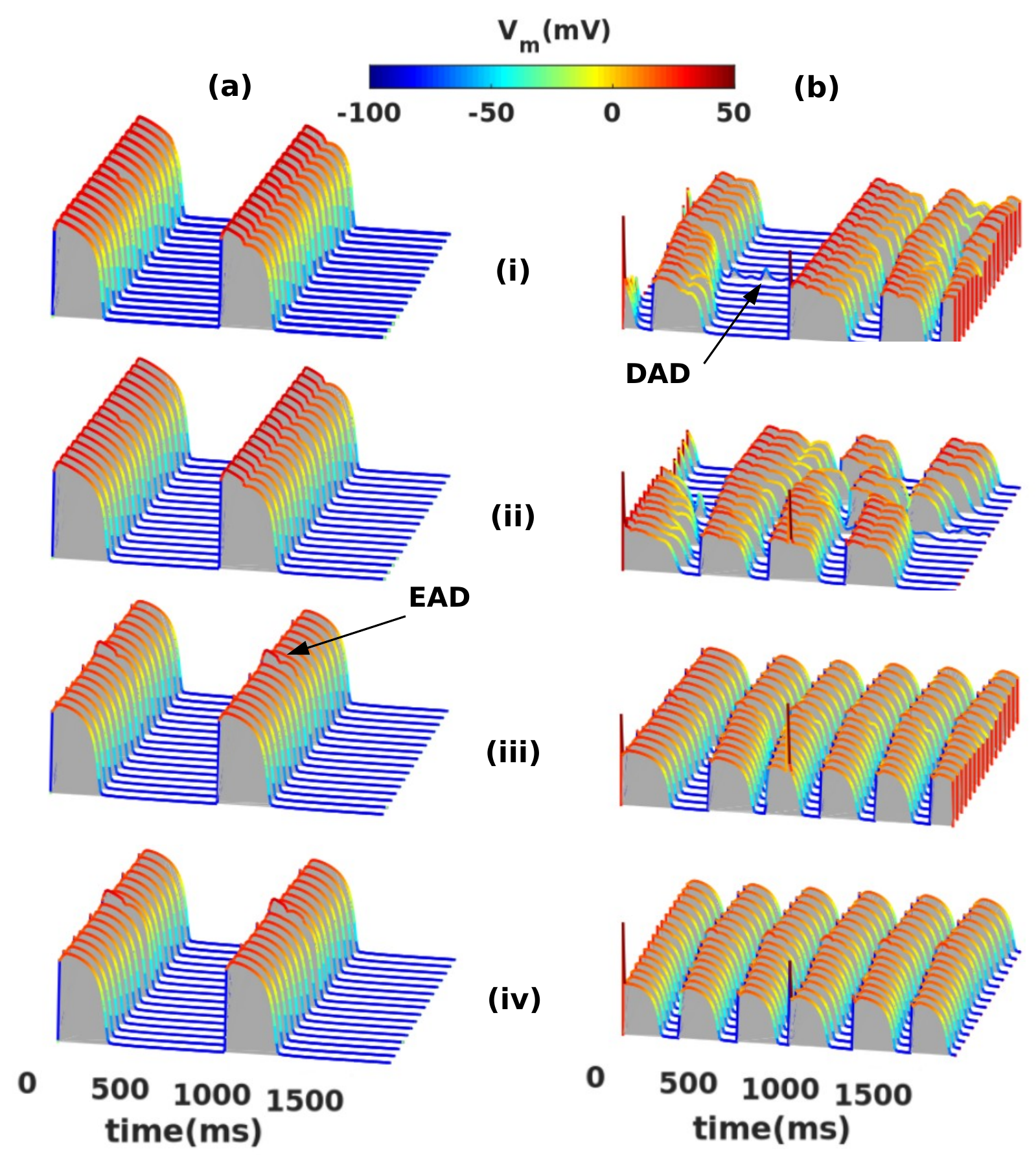}}
    \caption{(Color online) \label{fig:height_plot} Space-time plots (called waterfall plots in MATLAB) 
    of the membrane potential $V_{\rm{m}}$ (on the vertical axis), for the myocytes at sites $1, 16, 32, \ldots, 256$ along the cable, for Cases (i)-(iv) [see  Table~\ref{tab:cable_configs} and the text] for (a) the initial two pacings and the (b) after $120\,$s (the last two pacings). LCR-induced EADs and SCR-induced DADs are indicated at representative points by arrows.}
    \label{fig:cable}
\end{figure}

We present pseudocolor space-time plots of $V_{\rm{m}}$ for Cases (i)-(iv), in Fig.~\ref{fig:cable_cases} (b), for the initial two pacings, and Fig.~\ref{fig:cable_cases} (c), after $120$ pacings, with a pacing frequency of $1$ Hz. Initially, the stimulation, provided at one end of the cable, reaches the other end of the cable as in Figs.~\ref{fig:cable_cases}(b), (i)-(iv); however, after a few pacings, the $Ca_{\rm{SR}}$ load builds up in the myocytes and leads to suprathreshold DADs, which, in turn,  precipitate premature ventricular complexes (PVCs) in the cable; these PVCs overtake the periodic stimulation of the cable. The PVCs, which originate from the DAD clump, then stimulate the entire cable; however, after $120\,$s, signatures of conduction block appear in the cable for Case (ii); by contrast, in Cases (i), (iii), and (iv), we observe PVCs emerging from the center of the cable, from where they reach, uninterrupted, both the ends of the cable. 
In Figs.~\ref{fig:height_plot}(a) and (b), we present plots of the AP from the myocytes at sites $1, 16, 32, \ldots, 256$ along the cable to demonstrate the emergence of LCR-induced EADs and SCR-induced DADs (highlighted at representative points by arrows).

\begin{figure}
    \centerline{\includegraphics [width=\linewidth] {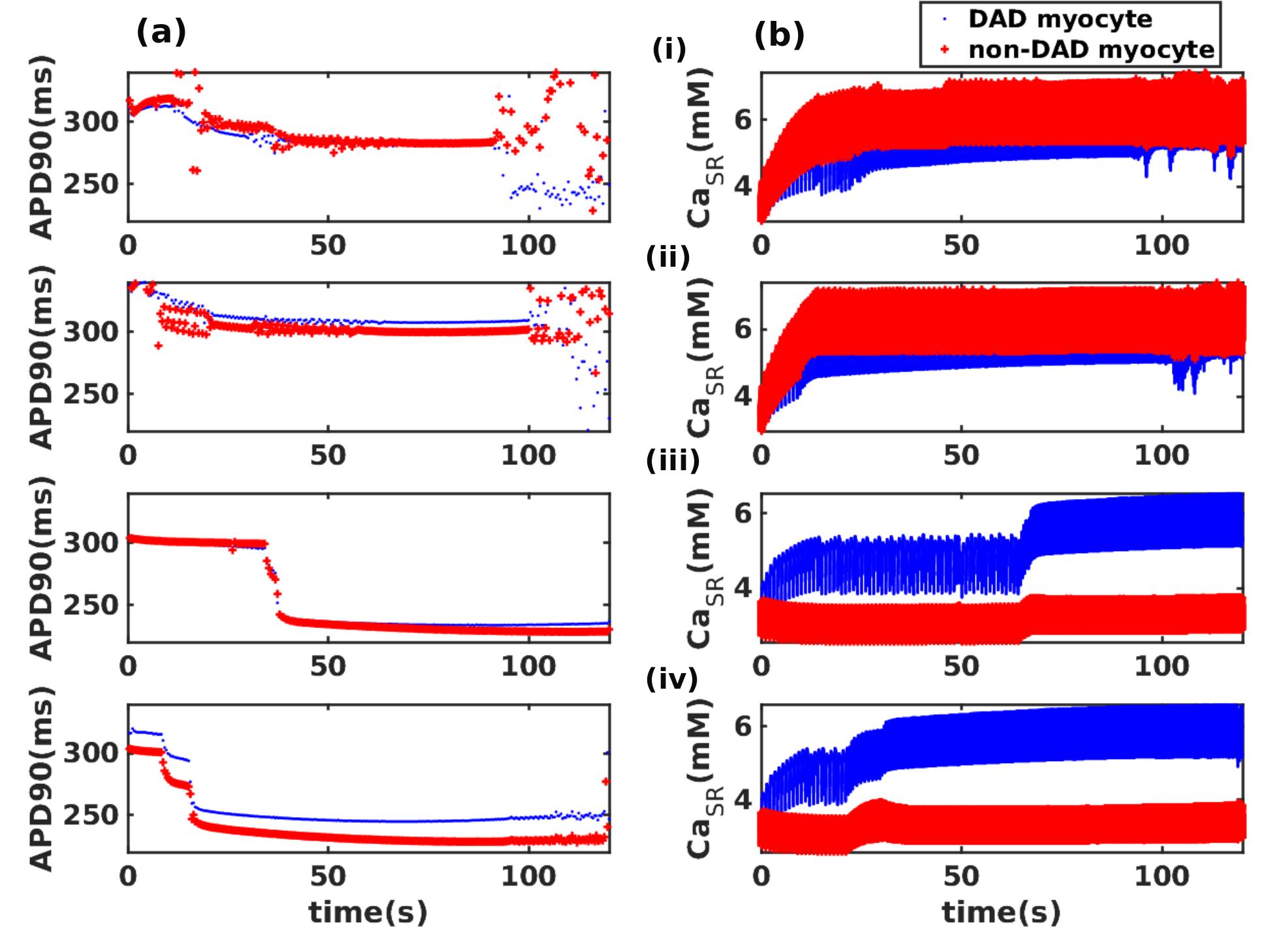}}
    \caption{(Color online) \label{fig:apd_casr} Plots versus time $t$ of (a) the APD and (b) $Ca_{\rm{SR}}$ for normal myocytes (blue points) and DAD-capable myocytes (red $+$) for Cases (i)-(iv) for the entire duration of our simulation in the cable domain of Fig.~\ref{fig:cable_cases} (a) from the representative points marked by red stars.}
\end{figure}

In Figs.~\ref{fig:apd_casr} (a) and (b), we present the temporal evolution of the APD and $Ca_{\rm{SR}}$, for non-DAD and DAD-capable myocytes from representative points [stars in Fig.~\ref{fig:cable_cases}(a)] in the simulation domain; for Cases (i)-(iv) [see Table~\ref{tab:cable_configs}] and for the entire $120\,$s duration of our cable simulation. The emergence of suprathreshold DADs, which have a higher frequency of incidence than the stimulation frequency, leads to the drop in the APD that is seen clearly in Figs.~\ref{fig:apd_casr}(a) (i)-(iv); this drop is significantly higher in Figs.~\ref{fig:apd_casr}(a) (iii)-(iv) than in Figs.~\ref{fig:apd_casr}(a) (i)-(ii).
After $100$s, the APDs for subsequent excitations are spread over a range, see Figs.~\ref{fig:apd_casr}(a) (i) and (ii)] but limited in Figs.~\ref{fig:apd_casr}(a) (iii) and (iv). Furthermore, Figs.~\ref{fig:apd_casr}(a) (i) and (ii) show that after $100\,$s the beat-to-beat APDs from DAD-capable and non-DAD myocytes can differ significantly, i.e., there is dispersion in the APD across these myocytes. Note that this difference in APDs, which is not present initially but develops as time progresses, can provide a substrate for reentry in 2D and 3D tissue (see below) ~\cite{clayton2005dispersion}.
For Cases (i) and (ii), i.e., for global Ca\textsuperscript{2+} overloads, the plots of $Ca_{\rm{SR}}$ versus time in Figs.~\ref{fig:apd_casr}(b) (i) and (ii)], for DAD-capable (blue points) and non-DAD myocytes (red $+$) overlap substantially
However, for localized Ca\textsuperscript{2+} overload, the plots of $Ca_{\rm{SR}}$ for the DAD-capable and non-DAD myocytes do not overlap [see Figs.~\ref{fig:apd_casr}(iii) and (iv)].

In summary, then, we have demonstrated how the calcium-overload-induced LCRs and SCRs, which appear at the single-cell level in DAD-capable myocytes [see Fig.~\ref{fig:pacing_effect} in Subsection \ref{subsec:afterdep}], can lead to dispersion in the APD and conduction blocks in the cable. Once the amplitude of these SCRs, from the myocytes in the DAD clump, is large enough to overcome the source-sink mismatch and go beyond the threshold for triggered excitations, DAD-driven PVCs emerge from this clump. The incidence frequency of these PVCs [which is much higher than the pacing frequency ($1$ Hz)] depends on the Ca\textsuperscript{2+} uptake rate in the SR (via the SERCA pump) and $Ca_{\rm{SR}}$. The PVC-driven high-frequency stimulation leads to more Ca\textsuperscript{2+} build-up in the SR and other myocyte compartments; this increased $Ca_{\rm{SR}}$ leads to an enhancement in the PVC frequency and results in a positive feedback loop between the PVC frequency and Ca\textsuperscript{2+}. The PVCs from the DAD clump stimulate the myocytes outside the clump, thereby increasing the extent of the Ca\textsuperscript{2+} overload; this promotes LCRs. After about $20\,$s, a significant difference emerges between the APDs from DAD and non-DAD myocytes and leads, in turn, to conduction block. We will show in the following Susections that such block leads to spiral- and scroll-wave formation in 2D and 3D tissue, respectively.

\begin{table*}[t]
	\centering 
	\begin{tabular}{|c|c|c|c|c|c|c|c|c|} 
	
	\hline
	 \multicolumn{1}{|c|}{\textbf{Configuration}} & \multicolumn{4}{c|}{{\textbf{DAD myocytes (RyR leak)}}} & \multicolumn{4}{c|}{{\textbf{non-DAD myocytes (no RyR leak)}}} \\
	\hline
	
Cases & {$S_{\rm{GCaL}}$}  & {$S_{\rm{GKr}}$} & {$S_{\rm{KNaCa}}$} & {$S_{\rm{Vmaxup}}$} & {$S_{\rm{GCaL}}$} & {$S_{\rm{GKr}}$} & {$S_{\rm{KNaCa}}$} & {$S_{\rm{Vmaxup}}$} \\ 
 \hline 
 \hline
  (i) & 2.5 & 2.0 & 2.0 & 4.5 & 2.5 & 2.0 & 2.0 & 4.5  \\
 \hline	
 
 (ii) & 2.5 & 1.0 & 2.0 & 4.5 & 2.5 & 1.0 & 2.0 & 4.5  \\
 \hline	

 (iii) & 2.5 & 2.0 & 2.0 & 4.5 & 1.0 & 1.0 & 1.0 & 1.0  \\
 \hline	
 
 (iv) & 2.5 & 1.0 & 2.0 & 4.5 & 1.0 & 1.0 & 1.0 & 1.0  \\
 \hline	
	
	\end{tabular}

	\caption{Scale factors used for different cases of cable simulations with DAD-capable and normal myocytes; in the DAD regions we consider cases with RyR leaks ($V_{\rm{RyRL}} = 0.00018\, {\rm ms^{-1}}$); myocytes in the non-DAD parts have zero RyR leak ($V_{\rm{RyRL}} = 0\, {\rm ms^{-1}}$).}
	\label{tab:cable_configs}
\end{table*}  
  
\subsection{From PVCs to spiral waves}
\label{subsec:2D}
We now generalize the study of Subsection~\ref{subsec:cable},
 of the effects of LCRs and SCRs, to a 2D simulation domain [and the TP06 model for ventricular tissue]. We concentrate on Case(ii) [Table~\ref{tab:cable_configs}], because it is most likely to cause conduction blocks in cable-type domains. We perform simulations in a rectangular domain with $512\times220$ grid points [$\simeq 12.5 \times 5.5$ cm$^2$]; we embed a clump of DAD-capable myocytes (henceforth DAD clump) with Case-(ii) parameters: for purposes of illustration, we employ either a circular clump (diameter $4$ cm or $160$ grid points) or a square clump (side $4$ cm or $160$ grid points).

\begin{figure*}
    \centerline{\includegraphics [width=\textwidth] {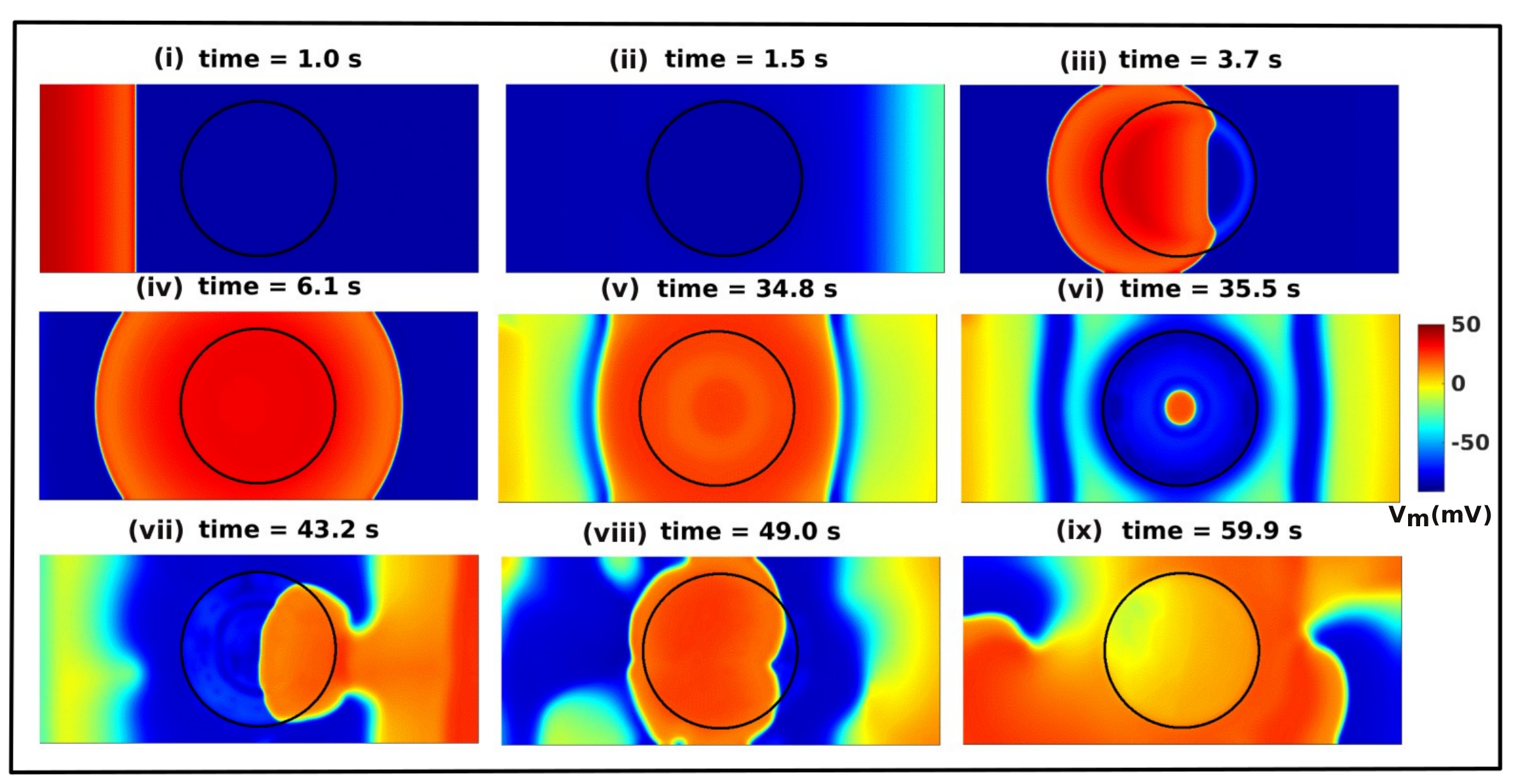}}
    \caption{(Color online)\label{fig:plane_to_spiral} Pseudocolor plots of the membrane potential $V_{\rm{m}}$ showing (i),(ii): propagation of plane waves; (iii),(iv): emergence and propagation of PVCs originating from a DAD clump; (v)-(viii): the collision of the wavefront and waveback of two successive PVCs; (ix) the formation of two stable spiral waves. For the complete spatiotemporal evolution of $V_{\rm{m}}$, see Video~\ref{S1_Video} in the Appendix.}
\end{figure*}

We stimulate the first two columns of myocytes at the left boundary of this domain, with a pacing frequency of $1$ Hz, and we apply no-flux boundary conditions on all the boundaries of the rectangle. We find that, after $3$ stimulations, PVCs emerge from such DAD clumps. With the passage of time, a wavefront, arising from a PVC that originates from the clump, meets the waveback of the previous PVC and results in a conduction block, which leads, in turn, to the development of rotating spiral waves (reentry) and, eventually, broken spiral waves (fibrillation):
In Fig.~\ref{fig:plane_to_spiral} we present pseudocolor plots of $V_{\rm{m}}$, at representative times $t$, to illustrate how plane-wave pacing [Figs.~\ref{fig:plane_to_spiral} (i) and (ii)] stimulates the DAD-clump and leads to PVCs [Figs.~\ref{fig:plane_to_spiral} (iii) and (iv)], which propagate outwards from this clump. The collisions of the wavefronts and wavebacks of two successive PVCs lead first to conduction blocks [Figs.~\ref{fig:plane_to_spiral} (v) and (vi)] and finally to the formation of spiral waves [two spiral-wave tips appear in Figs.\ref{fig:plane_to_spiral} (ix)]. For the complete spatiotemporal evolution of $V_{\rm{m}}$, see Video~\ref{S1_Video} in the Appendix.\\
In Fig.~\ref{fig:apd_evol_2d}, we present the time series of the APDs extracted from plots of $V_{\rm{m}}(t)$, from the two representative points, one outside the DAD clump (i) and the other inside it (ii) [see Fig.~\ref{fig:apd_evol_2d}(a)]; here, Fig.~\ref{fig:apd_evol_2d}(b)(i) is the plot of APDs, obtained from the non-DAD region, and Fig.~\ref{fig:apd_evol_2d}(b)(ii) is the APD from the DAD region. As time progresses, the APD of the myocytes in the DAD region is stable; however, the APD from the non-DAD myocytes varies in time. The transition to reentry, after the conduction block, is reflected in the sudden jump in the APD of the DAD myocytes and the large variations in the APD of the non-DAD myocytes.

\begin{figure}
    \centerline{\includegraphics [width=0.5\textwidth] {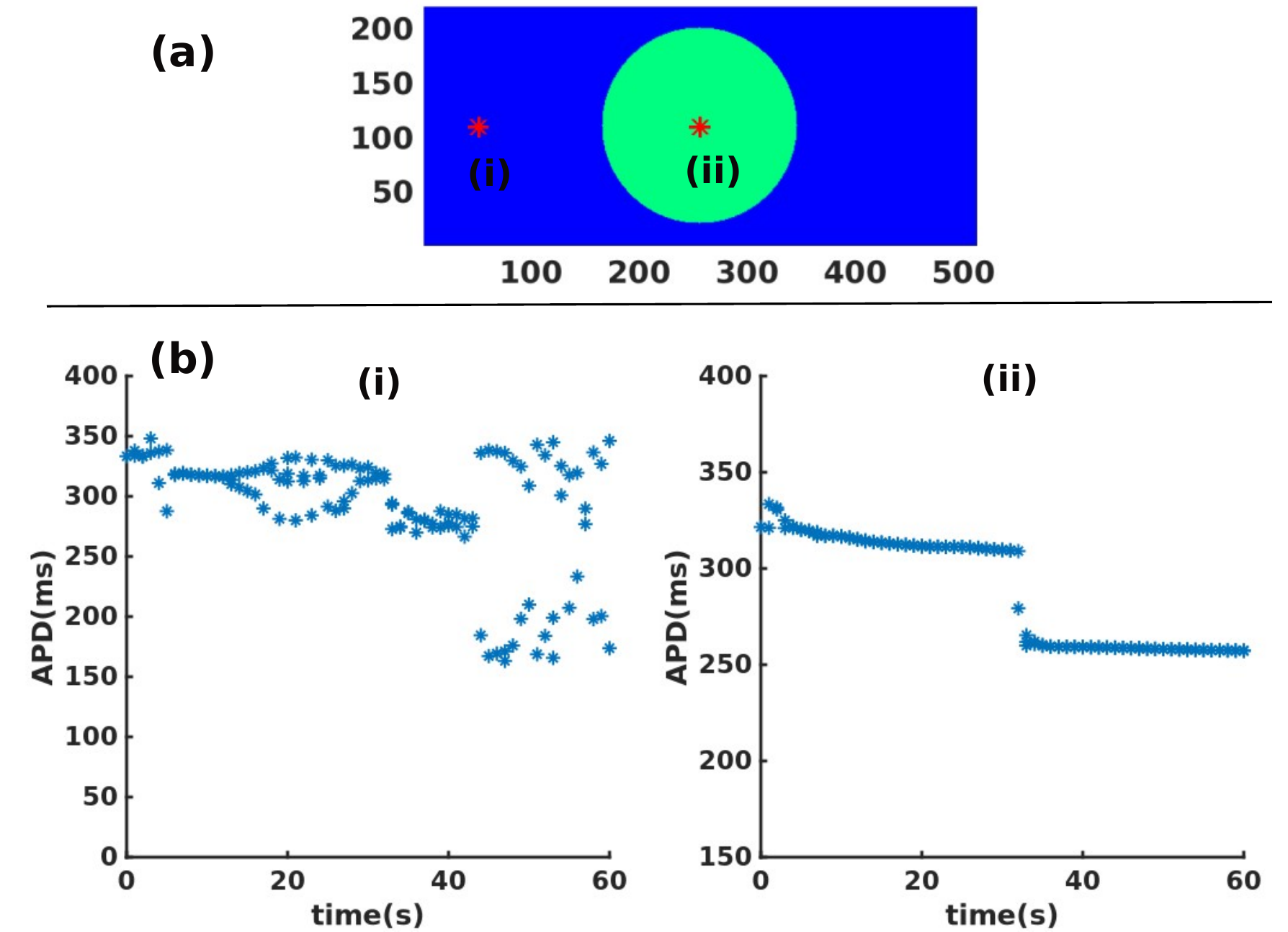}}
    \caption{ (Color online) \label{fig:apd_evol_2d} The evolution of the APDs, from $V_{\rm m}(t)$ at representative points (shown in red in (a)) inside the DAD clump (green) and the normal-myocyte region (blue) in a rectangular domain. (b) APDs of the subsequent excitations from (i) myocytes outside the DAD clump and (ii) in the middle of this clump.}
\end{figure}

\begin{figure}
    \centerline{\includegraphics [width=0.5\textwidth] {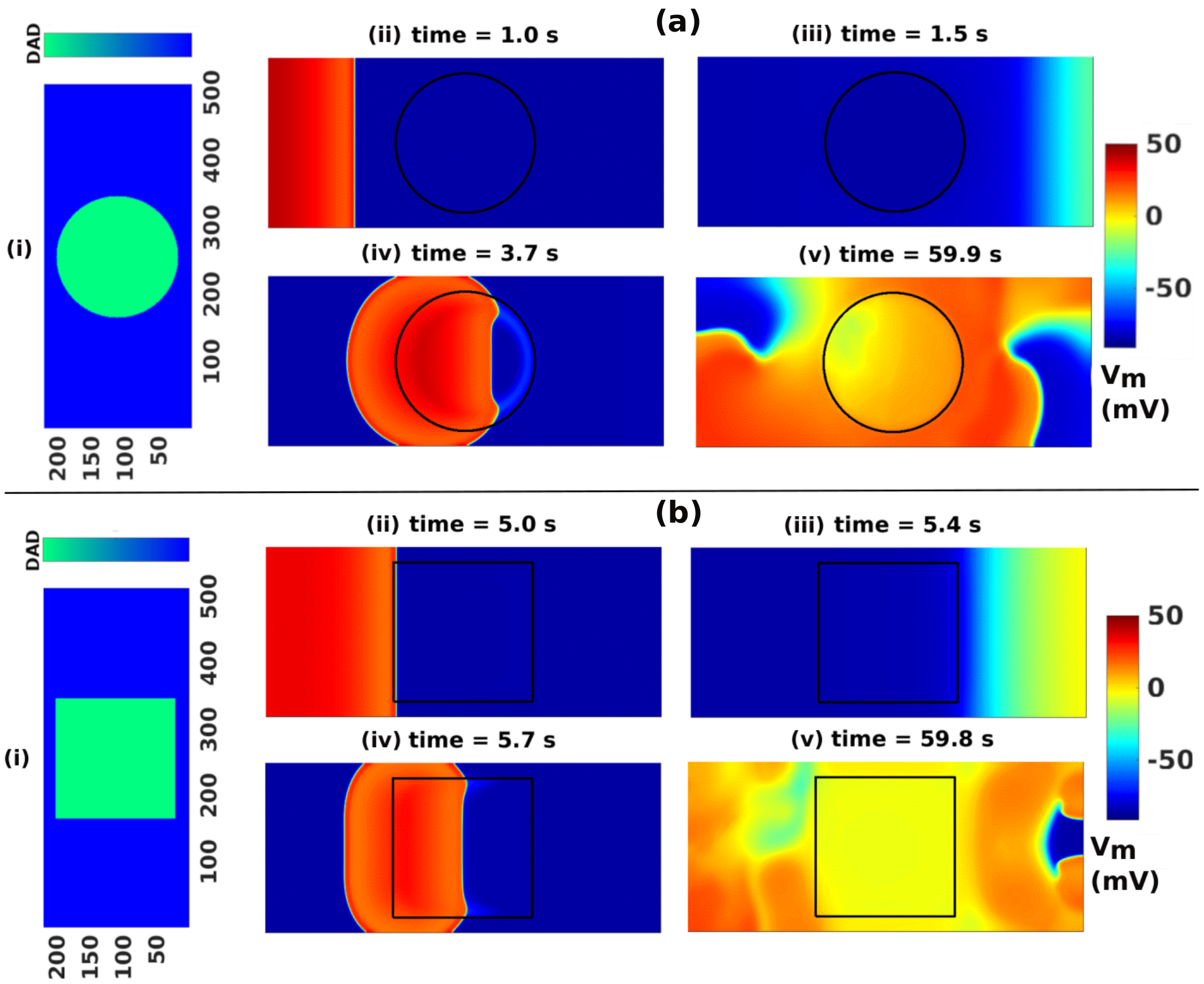}}
    \caption{ (Color online) \label{fig:geom_dep} Pseudocolor plots comparing the effects of (a) circular [with diameter $= 160$ grid points] and (b) square [with side $= 160$ grid points] DAD clumps, with all other model parameters held fixed. In each of the subplots (a) and (b): (i) shows pseudocolor plots of the domain, containing non-DAD (blue) and DAD (green) myocytes; (ii),(iii): pseudocolor plots of  $V_{\rm{m}}$ showing pacing-induced plane-wave propagation; (iv): the emergence of PVCs; (v): 
    the evolution of PVCs into reentrant spiral waves. For the complete spatiotemporal evolution of $V_{\rm{m}}$ for these cases see Video ~\ref{S1_Video} and Video ~\ref{S2_Video} in the Appendix.}
\end{figure}

In Fig.~\ref{fig:geom_dep} we compare the effects of circular [with diameter $= 160$ grid points] and square [with side $= 160$ grid points] DAD clumps, with all other model parameters held fixed. If we stimulate the first two columns of myocytes at the left boundary of this domain, with a pacing frequency of $1$ Hz, PVCs originate from both these clumps. The PVC from the square clump is flatter in parts than the one from the circular clump, as we can see by comparing Figs.~\ref{fig:geom_dep}(a)(iv) and (b)(iv). As time progresses, conduction block occurs [see, e.g., Figs.~\ref{fig:plane_to_spiral} (v) and (vi) for the circular clump]; and then two spiral-wave cores form, more clearly for the circular clump than for the square one. [For the complete spatiotemporal evolution of $V_{\rm{m}}$ for these cases see Video ~\ref{S1_Video} and Video ~\ref{S2_Video} in the Appendix.] We conjecture that this difference arises because of the flatness of the PVCs from the rectangular clump. \\

We now use the low-amplitude defibrillation scheme suggested in Refs.~\cite{sinha2001defibrillation,shajahan2009spiral}, in
which we divide the simulation domain into square subdomains of size $30\times30$ grid points; at the boundaries of each subdomain, we apply a current stimulus (to each one of the $3$ grid points that straddle every point on these boundaries) with amplitude $-50$ pA/pF for $10$ ms. Such a current stimulus, applied on the spatially extended mesh formed by the boundaries of the subdomains, yields the mathematical analog of defibrillation, i.e., spiral waves, generated by using the S1-S2 cross-field protocol (see, e.g., Ref.~\cite{zimik2016instability}) in the original TP06 model~\cite{ten2006alternans}, are eliminated, as we show by representative pseudocolor plots of $V_{\rm{m}}$ in Fig.~\ref{fig:defib}(a) and Video ~\ref{S3_Video} in the Appendix. We find, however, that this defibrillation fails to eliminate PVCs that originate from the DAD clump that we have studied above; we illustrate this by representative pseudocolor plots of $V_{\rm{m}}$ in Fig.~\ref{fig:defib}(b) and Video~\ref{S4_Video} in the Appendix. Our results are reminiscent of those reported in Refs.~\cite{vandersickel2014study,zimik2015comparative} for models with EAD-capable myocytes in the whole simulation domain; these models display, \textit{inter alia}, phase waves that move through the types of meshes we have used above. Therefore, one way of understanding the failure of our defibrillation scheme in Fig.~\ref{fig:defib}(b) is to realise that phase waves form inside the DAD clump, pass through it, unimpeded by the current stimulus on the mesh, and continue to yield PVCs. Defibrillation failure in such arrhythmias occurs because, at the wavefront, the current $I_{\rm{NaCa}}$ leads $I_{\rm{CaL}}$ and $I_{\rm{Na}}$, as we show explicitly via the pseudocolor plots of $V_{\rm{m}}$ and these currents in Figs.~\ref{fig:lead} (a)-(d) and the Video~\ref{S5_Video} in the Appendix. 
Such PVC-induced arrhythmias cannot be eliminated by low-amplitude current stimuli in our model [see Fig.~\ref{fig:defib} (b)]. 
By contrast, in conventional arrhythmias, with self-sustaining spirals, $I_{\rm{Na}}$ is the lead current, which can be eliminated by the application of small current stimuli for a brief duration [as we show in Fig.~\ref{fig:defib} (a)].

\begin{figure*}
    \centerline{\includegraphics [width=\textwidth] {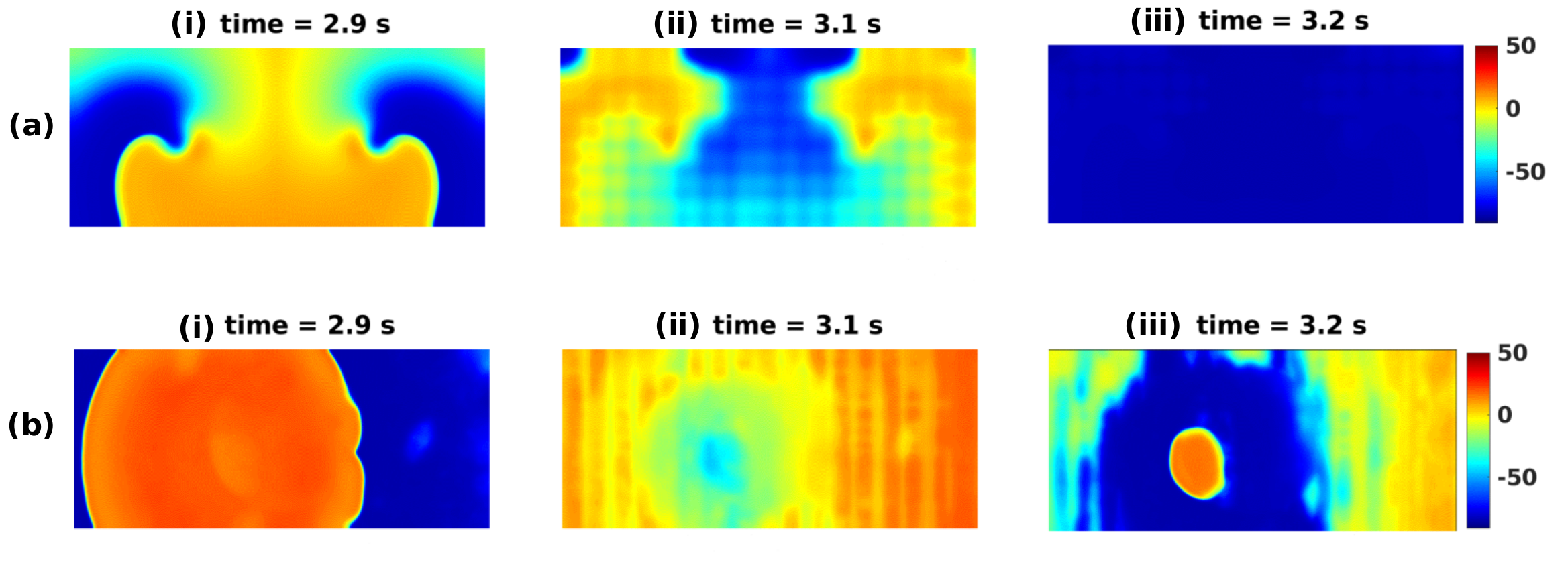}}
    \caption{ (Color online) \label{fig:defib} Pseudocolor plots of the membrane potential $V_{\rm{m}}$ showing the elimination of arrhythmogenic spiral waves in (a) the conventional 2D TP06 model for cardiac tissue and (b) this model with a DAD clump. (a)[(i)-(iii)]: the termination of spiral waves by using the low-amplitude defibrillation scheme suggested in Refs.~\cite{sinha2001defibrillation,shajahan2009spiral}. (b)[(i)-(iii)]: This defibrillation scheme is unsuccessful in terminating PVCs that originate from the DAD clump. For the complete spatiotemporal evolution of $V_{\rm{m}}$, see Video ~\ref{S3_Video} and Video~\ref{S4_Video} in the Appendix.}
\end{figure*}

\subsection{Human bi-ventricular simulation}
\label{subsec:3D}
\begin{figure}
    \centerline{\includegraphics [width=8.5cm] {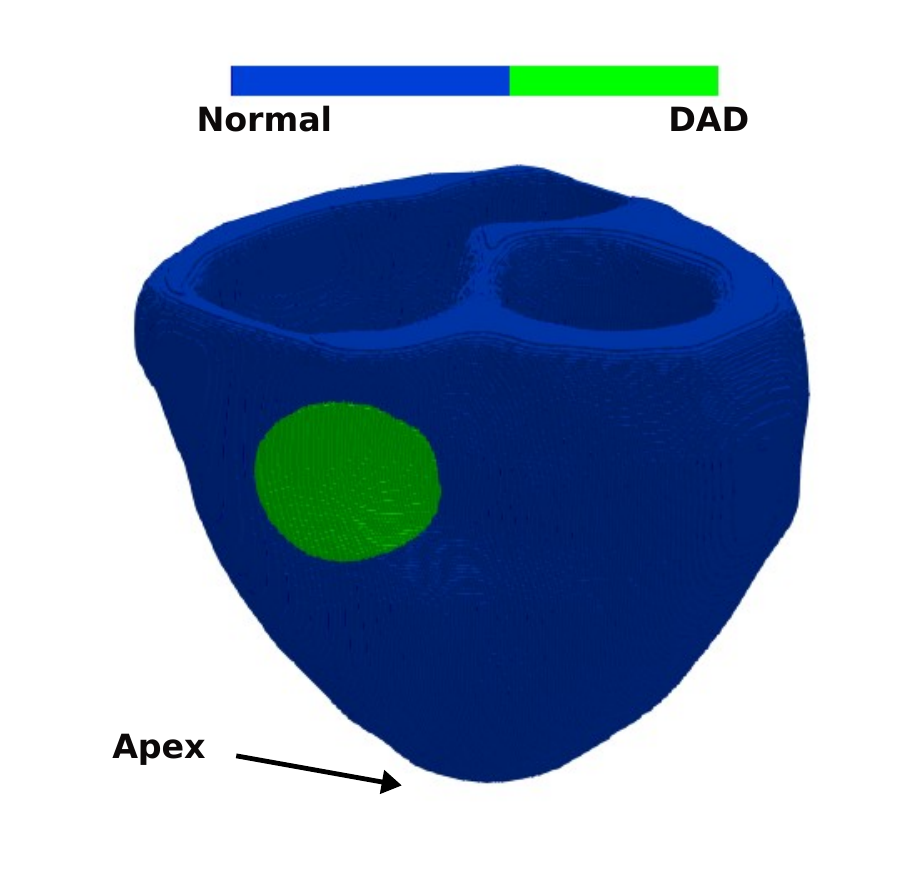}}
    \caption{ (Color online) \label{fig:DAD_heart} A pseudocolor plot showing a representative 3d human bi-ventricular geometry with a DAD clump.}
\end{figure}

\begin{figure*}
    \centerline{\includegraphics [width=\textwidth] {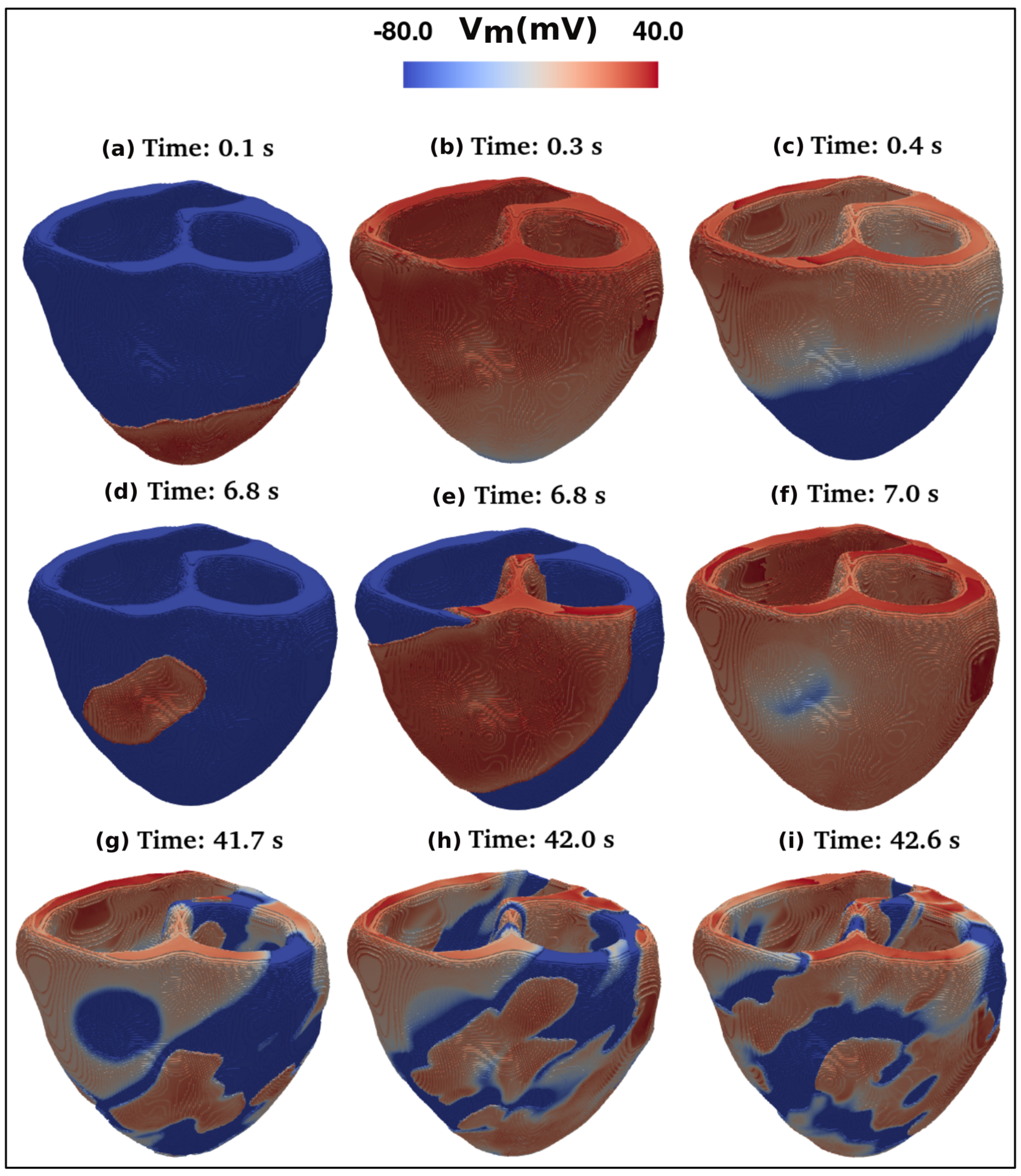}}
    \caption{(Color online) \label{fig:3d} Pseudocolor plots of $V_{\rm{m}}$, superimposed on the bi-ventricular simulation domain (in blue), depicting the spatiotemporal evolution of the electrical excitations, the emergence of PVCs, and the evolution of PVCs to scroll and broken scroll-waves. For the
complete spatiotemporal evolution of $V_{\rm{m}}$ for this case see Video~\ref{S5_Video} in the Appendix.}
\end{figure*}

Human ventricular tissue is anisotropic because of the orientation of muscle fibers; furthermore, the anatomically realistic human bi-ventricular domain is complicated. We use the human bi-ventricular geometry [obtained from diffusion tensor magnetic resonance imaging (DTMRI)], which we enclose in a cubical box with $512\times512\times512$ grid points. The bi-ventricular domain inside this cube is conveniently described by using the phase-field variable $\phi$, with $\phi=1$ in this domain and $\phi=0$ outside [see, e.g., Refs.~\cite{rajany2022spiral,majumder2016scroll,winslow2011cardiovascular,fenton2005modeling}]; $\phi$ changes continuously from $1$ to $0$, typically over $4-5$ grid points. We take the DAD clump to be the region of overlap of this bi-ventricular geometry with a sphere; this clump is the green region (with $775,596$ grid points) in the pseudocolor plot of Fig.~\ref{fig:DAD_heart}, in which the region with non-DAD myocytes is shaded blue. To generate the Ca\textsuperscript{2+}-overload for the DAD-capable and non-DAD myocytes, we use Case-(ii) parameters [Table~\ref{tab:cable_configs}]. We pace  $40\times512\times512$ myocytes at the apex of the bi-ventricular domain [Fig.~\ref{fig:DAD_heart}]. The pseudocolor plot of the membrane potential of $V_{\rm{m}}$, in Figs.~\ref{fig:3d} (a)-(i), shows how the initial, pacing-induced plane wave [Fig.~\ref{fig:3d} (a)] interacts with the DAD clump to yield PVCs, after the initial five pacings [Fig.~\ref{fig:3d}(d)]; these PVCs stimulate the entire bi-ventricular domain, with a frequency higher than the $1$ Hz pacing frequency provided at the apex. These PVCs evolve into a rotating scroll wave that eventually becomes a broken scroll wave [see Figs.~\ref{fig:3d}(g)-(i) and the Video~\ref{S5_Video} in the Appendix and compare these with our 2D-tissue results in Subsection~\ref{subsec:2D}].

\section{Discussion and Conclusions}
\label{sec:disc_and_concl}

\begin{figure*}
    \centerline{\includegraphics [width=\textwidth] {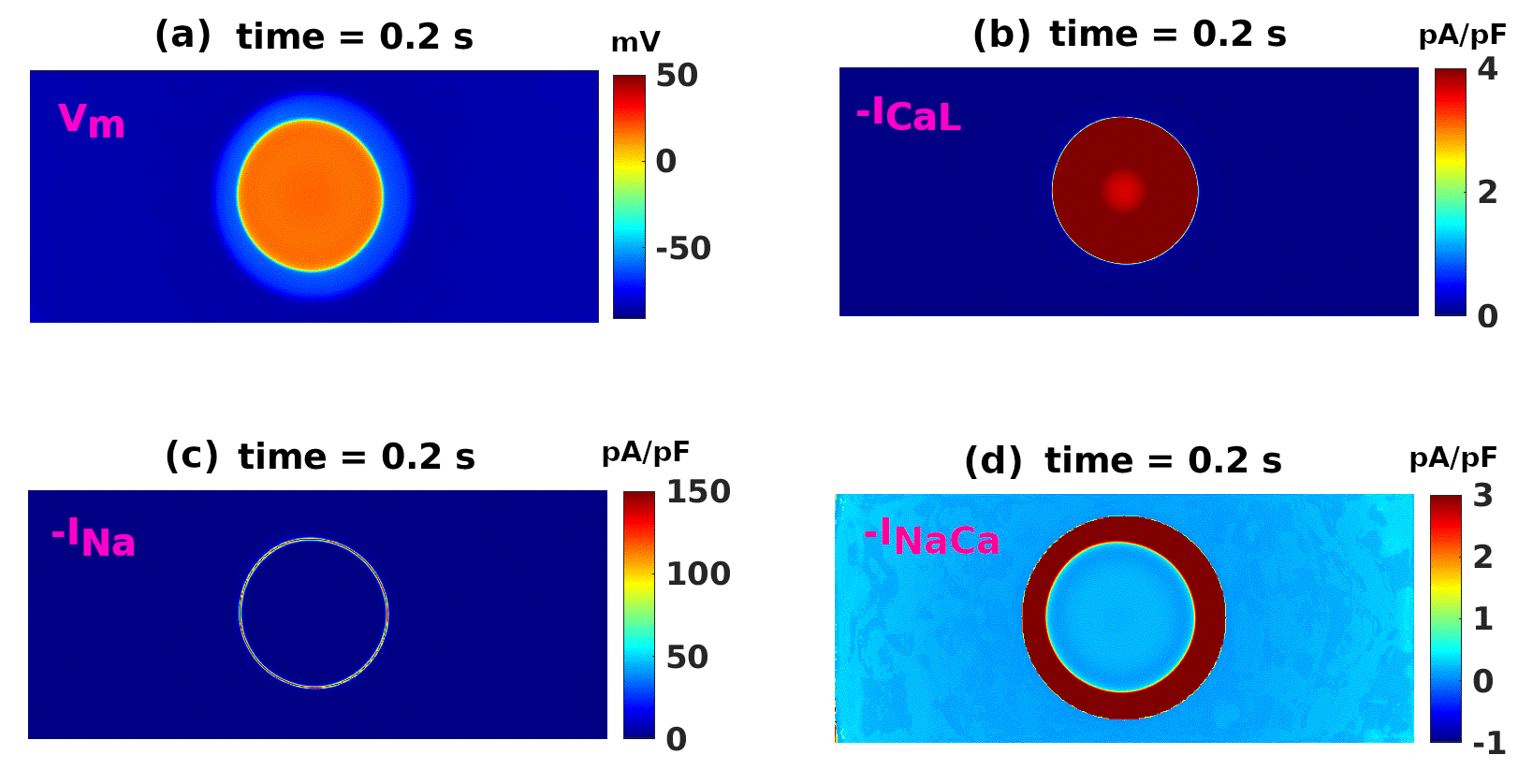}}
    \vspace{1cm}
    \caption{(Color online) \label{fig:lead} Pseudocolor plots of the membrane potential and a few currents during the emergence of PVC, (a) $V_{\rm{m}}$; (b) $-I_{\rm{CaL}}$;  (c) $-I_{\rm{Na}}$; and (d) $-I_{\rm{NaCa}}$. For the complete spatiotemporal evolutions see Video~\ref{S6_Video} in the Appendix.}
\end{figure*}

We have presented a detailed multiscale study -- from a single myocyte  to an anatomically realistic bi-ventricular domain -- of LCR-induced EADs, SCR-triggered DADs, and DAD-clump-promoted PVCs in the TP06 mathematical model for human ventricular tissue. We first show [Subsection~\ref{subsec:protocol}] how an increase of $I_{\rm{CaL}}$ provides the Ca\textsuperscript{2+}-overload and the way in which this affects $V_{\rm{m}}$ and the Ca\textsuperscript{2+} concentration in the sarcoplasmic store, for two Ca\textsuperscript{2+}-overload protocols. We demonstrate [Subsection~\ref{subsec:afterdep}] the emergence of LCR-induced EADs and SCR-triggered DADs and their dependence on the RyR. Next, we explore [Subsection~\ref{subsec:lcclate}] how a selective reduction of the late part of $I_{\rm{CaL}}$ can eliminate LCRs and hence EADs; we distinguish between DAD myocytes and non-DAD myocytes. 
This leads naturally to our investigation of the propagation of electrical waves of activation in a cable-type domain [Subsection~\ref{subsec:cable}] with non-DAD myocytes and a clump of DAD myocytes in the middle. By examining this propagation for the two Ca\textsuperscript{2+}-overload protocols [Subsection~\ref{subsec:protocol}], with the Ca\textsuperscript{2+}-overload either localized to the DAD-clump or spread over the entire cable. In the latter case, we demonstrate the development of DAD-clump-driven PVCs, and in turn, to conduction blocks. Our simulations in two-dimensional (2D) [Subsection~\ref{subsec:2D}] and three-dimensional (3D) anatomically realistic human bi-ventricular [Subsection~\ref{subsec:3D}] domains, with DAD clumps, also display such PVCs that lead to reentry and the formation of spiral (2D) or scroll (3D) waves. For a recent overview of all types of PVCs and their clinical implications and management, we refer the reader to Ref.~\cite{marcus2020evaluation}; this paper notes that PVCs occur often and they " \ldots are observed in the majority of individuals monitored for more than a few hours \ldots ". We have studied those PVCs that arise from DAD clumps; and we have examined the conditions under which these PVCs evolve into life-threatening ventricular fibrillation. 
Earlier studies have also investigated other LCR- and SCR-induced DADs and EADs. For example, the authors of Ref.~\cite{volders1997similarities} have used isoproterenol-treated myocytes to generate such afterdepolarizations; they have then discussed the possibility of the dispersion of APDs, among EAD myocytes, as a substrate for reentry. Myocytes from different transmural layers of ventricles may have different APD elongation, in response to $\beta$-adrenergic stimulation, which can be another possible mechanism for reentry and fibrillation~\cite{volders1997similarities, antzelevitch1995clinical}. Reference~\cite{shiferaw2012intracellular} has suggested, based on experiments~\cite{wasserstrom2010variability} and a single-myocyte model~\cite{chen2011mathematical}, that triggered waves associated with DADs and EADs might lead to ectopic foci and thence to arrhythmias. We have explored this last suggestion via detailed multiscale simulations, by using the TP06 model for human ventricular tissue.\\
We note the following points: (1) Although our study is based on a ventricular-myocyte model, it can be generalized \textit{mutatis mutandis} to study 
LCRs and SCRs in atrial tissue and to elucidate their roles in atrial arrhythmias [see, e.g., see Ref~\cite{hove2004atrial}].
(2) We have used Ca\textsuperscript{2+}-overload to trigger SCR and LCRs; however, LCRs and SCRs can occur without Ca\textsuperscript{2+}-overload, e.g., if there is heart failure. Irrespective of the way in which LCRs and SCRs are generated, they should lead to PVCs and the formation of spiral (2D) or scroll (3D) waves as we have discussed above.\\
We hope that our detailed study of DAD-clump-promoted PVCs and fibrillation  will lead to experimental verifications of our results, e.g., in ~\textit{in vitro} experiments with engineered human-heart tissue obtained from pluripotent stem cells [see, e.g., Ref.~\cite{park2019insights}].\\

\subsection{Possible clinical implications of our study}
\label{subsec:Clinical}

Disturbed Ca\textsuperscript{2+} homeostasis is intimately associated with cardiac arrhythmias as discussed, e.g., in Refs.~\cite{deo2017calcium,landstrom2017calcium,pogwizd2004cellular}. The disruption in Ca\textsuperscript{2+} homeostasis has been observed in a variety of conditions, e.g.,  heart failure (HF)~\cite{fowler2020arrhythmogenic} and catecholaminergic-polymorphic ventricular tachycardia (CPVT)~\cite{wleklinski2020molecular}. Reference~\cite{shiferaw2012intracellular} has suggested, based on experiments~\cite{wasserstrom2010variability} and a single-myocyte model~\cite{chen2011mathematical}, that triggered waves associated with DADs and EADs might lead to ectopic foci and thence to arrhythmias, a possibility that we have examined in detail for the TP06 model. In particular, we have considered two Ca\textsuperscript{2+}-overload protocols, one with an increased APD and the other without an increase in the APD; the former protocol, with Ca\textsuperscript{2+} overload in the entire domain, leads to conduction block, reentry, and fibrillation. Our study demonstrates (a) the protective mechanism of the repolarizing current $I_{\rm{Kr}}$, in the development of these
Ca\textsuperscript{2+}-overload-induced arrhythmias and (b) that the selective reduction of $I_{\rm{CaL}}$, in the plateau phase of the AP, can suppress LCRs, in agreement with the suggestions of Refs.~\cite{fowler2018late,fowler2020arrhythmogenic}; both these possibilities (a) and (b) can be used in the development of drugs for the suppression of such arrhythmias. Our study suggests [see Subsection~\ref{subsec:2D}] 
that it might not be easy to eliminate these arrhythmias by electrical defibrillation.
 \\
We have noted, while discussing Fig.~\ref{fig:f_sat}, that the reduction in $f_{\rm{2}{,sat}}$ reduces the magnitude of $I_{\rm{CaL}}$ (subplot (ii)), in its plateau phase (but not in the initial transient), suppression of oscillations in $Ca_{\rm{SR}}$ (subplot (iii)), and the elimination of the blue spikes in $I_{\rm{rel}}$ (subplot (iv)).

Therefore, medications that selectively target the plateau phase of $I_{\rm{CaL}}$ can eliminate LCRs, without affecting the transient phase of $I_{\rm{CaL}}$. References~\cite{frommeyer2012kalzium,laurita2008mechanisms} suggest that blocking
$I_{\rm{CaL}}$-related channels can eliminate LCR-induced arrhythmias; our study yields a more nuanced suggestion, to wit,
we must suppress only the plateau phase of $I_{\rm{CaL}}$ to remove LCRs. We observe that an increase of $I_{\rm{CaL}}$ or a reduction of $I_{\rm{Kr}}$ or both can lead to an enhancement of the APD [see Fig.~\ref{fig:f_sat}]. However, from Figs.~\ref{fig:protocol}, \ref{fig:cable_cases}, and \ref{fig:cable}, we conclude that, if we strengthen $I_{\rm{Kr}}$, by increasing $S_{\rm{GKr}}$, we can control the APD and, therefore, suppress reentry and fibrillation.\\

\subsection{Limitations of our study}

We have not used sub-cellular, microscopic descriptions for SCRs and LCRs. The synchronization of the SCRs and DADs across the myocyte is related, in our model, to the previous AP; however, the synchronization process is more complicated, if we use microscopic descriptions of the SCRs and LCRs. 
We have induced Ca\textsuperscript{2+} overload principally by controlling $I_{\rm{CaL}}$. We have not included overload via the stimulation of $\beta$-adrenergic receptors ($\beta$-AR) [see, e.g., Refs.~\cite{kimura1984delayed,myles2012local}]; heterogeneous $\beta$-AR stimulation can generate a spatially heterogeneous APD distribution, which can, in turn, act as a substrate for reentry; such stimulation enhances the L-type calcium channel (LCC) in the cardiac myocyte via the pathway of intracellular cyclic AMP (cAMP) and protein kinase A (PKA)~\cite{heijman2011local, saucerman2003modeling,ganesan2006beta}; this can, in turn, enlarge the APD
and thus increase the Ca\textsuperscript{2+}-overload and abnormal calcium releases as discussed,
e.g., in the models of Refs.~\cite{doste2021multiscale,saucerman2003modeling}. We note that Ca\textsuperscript{2+}-overload-induced afterdepolarizations and subsequent PVCs can occur not only in ventricular myocytes, but also for myocytes in the atria, Purkinje fibers, and pacemaker cells [see, e.g., Refs.~\cite{verkerk2000calcium,shah2019delayed,joung2011delayed}];  these are not included in our study. We have not accounted for electrical heterogeneity across the ventricular wall~\cite{antzelevitch1995clinical,antzelevitch2001electrical} and in the apico-basal direction~\cite{okada2011transmural}.\\
\section*{Data and code availability}

Data from this study and the computer scripts can be obtained from the authors upon 
reasonable request.

\section*{Conflicts of Interest}
No conflicts of interests, financial or otherwise, are declared by the authors.

\section*{Author Contributions} NR and RP planned the research and analysed the numerical data; NR carried out the calculations and prepared the tables, figures, and the draft of the manuscript; NR and RP then revised the manuscript in detail and approved the final version.

\section*{Funding}
We thank the Science and Engineering Research Board (SERB) and Council for Scientific and Industrial
Research (CSIR), and the National Supercomputing Mission (NSM), India for support,  and the Supercomputer Education and Research Centre (IISc) for computational resources.
\acknowledgments
We thank Mahesh K. Mulimani and Soling Zimik for valuable discussions.  

\appendix
\section*{Appendix}
\begin{widetext}

\subsection{Video SV1}
\label{S1_Video}
Animations of pseudocolor plots of the membrane potential $V_{\rm{m}}$ showing the propagation of plane waves, the emergence and propagation of PVCs originating from a DAD clump, the collision of the wavefront and waveback of two successive PVCs, and the formation of two stable spiral waves as in Fig.~\ref{fig:plane_to_spiral}. For all the videos here, we use $30$ frames per second, with each frame separated from the succeeding frame by $20$ms in real time. 
See video here: \url{https://youtu.be/MDqVnDQ-8sg}.

\subsection{Video SV2}
\label{S2_Video}
Animations of pseudocolor plots of the membrane potential $V_{\rm{m}}$ showing the propagation of plane waves, the emergence and propagation of PVCs originating from a DAD clump, the collision of the wavefront and waveback of two successive PVCs, and the formation of two spiral cores (temporarily) for a square DAD clump as in Fig.~\ref{fig:geom_dep}. See video here: \url{https://youtu.be/ptcnRiMi9iA}.

\subsection{Video SV3}
\label{S3_Video}

Animations of pseudocolor plots of the membrane potential $V_{\rm{m}}$ showing the elimination of arrhythmogenic spiral waves in the conventional 2D TP06 model for cardiac tissue by using the low-amplitude defibrillation scheme suggested Refs.~\cite{sinha2001defibrillation,shajahan2009spiral} (cf., Fig~\ref{fig:defib}). See video here: \url{https://youtu.be/x8aBZsdNZMc}.

\subsection{Video SV4}
\label{S4_Video}

Pseudocolor plots of the membrane potential $V_{\rm{m}}$ showing the elimination of arrhythmogenic PVCs in the TP06 model for cardiac tissue with a DAD clump by using the low-amplitude defibrillation scheme suggested in Refs.~\cite{sinha2001defibrillation,shajahan2009spiral}
(cf., Fig~\ref{fig:defib}). See video here: \url{https://youtu.be/GWpFT0kD__Q}.

\subsection{Video SV5}
\label{S5_Video}

Animations of pseudocolor plots of $V_{\rm{m}}$, superimposed on the bi-ventricular simulation domain (in blue), depicting the spatiotemporal evolution of the electrical excitations, the emergence of PVCs, and the evolution of PVCs to scroll and broken scroll-waves (cf., Fig~\ref{fig:3d}). See video here: \url{https://youtu.be/BAtsmGVwiQE}.

\subsection{Video SV6}
\label{S6_Video}

Animations of pseudocolor plots of the membrane potential and a few currents during the emergence of PVC, (top left) $V_{\rm{m}}$, (top right) $-I_{\rm{CaL}}$,  (bottom left) $-I_{\rm{Na}}$, and (bottom right) $-I_{\rm{NaCa}}$ (cf., Fig~\ref{fig:lead}). See video here: \url{https://youtu.be/B44qx-kaNA4}.

\end{widetext}

\bibliography{main}

\end{document}